\documentclass[aps,prl,twocolumn,superscriptaddress,nofootinbib]{revtex4-2}
\usepackage{amsmath,amssymb,graphicx}
\usepackage[colorlinks=true,allcolors=blue]{hyperref}
\usepackage[mathlines, switch]{lineno}
\begin{document}

\title{Geometric realization of stress-tensor deformed field theory}
\author{Yun-Ze Li}
\email{lyz21@mails.jlu.edu.cn}
\affiliation{Center for Theoretical Physics and College of Physics, Jilin University, Changchun 130012, People's Republic of China}

\author{Yunfei Xie}
\email{jieyf22@mails.jlu.edu.cn}
\affiliation{Center for Theoretical Physics and College of Physics, Jilin University, Changchun 130012, People's Republic of China}

\author{Song He}
\email{hesong@nbu.edu.cn}
\thanks{Corresponding author.}
\affiliation{Institute of Fundamental Physics and Quantum Technology \& School of Physical Science and Technology, Ningbo University, Ningbo, Zhejiang 315211, China}
\affiliation{Center for Theoretical Physics and College of Physics, Jilin University, Changchun 130012, People's Republic of China}
\affiliation{Max Planck Institute for Gravitational Physics (Albert Einstein Institute), Am M\"uhlenberg 1, 14476 Golm, Germany}

\begin{abstract}

We present a semiclassical framework in which stress--tensor deformations of a quantum field theory (QFT) reorganize into a gravitational action evaluated at a metric saddle. The deformed partition function can be written as a gravitational path integral evaluated at the saddle, establishing a direct link between stress--tensor flows and gravitational dynamics. Two complementary routes arise: (i) from gravitational actions such as Einstein and Palatini, which map to stress--tensor deformations of a seed QFT; and (ii) from deformed QFTs such as generalized Nambu--Goto and $T\bar{T}$-like deformed models, which reconstruct the corresponding gravitational actions. Finally, in a free, massive scalar theory, we show that the one-loop effective action of the nonlocal deformation contains a local curvature term; its coefficient defines an induced Newton constant at a chosen renormalization scale, thereby demonstrating a bidirectional link between stress--tensor flows and classical gravity.

\end{abstract}

\maketitle

\section{Introduction}
Understanding whether and how spacetime geometry can be encoded in quantum field theory (QFT) remains a central problem in theoretical physics~\cite{Visser:2002ew}. Thermodynamic arguments suggest that Einstein’s equations may reflect equilibrium conditions of quantum matter~\cite{Jacobson:1995ab, Padmanabhan:2010xh}, while holographic dualities such as AdS/CFT~\cite{Maldacena:1997re, Witten:1998qj} provide a framework for reconstructing spacetime from large-$N$ QFT data under specific boundary conditions, e.g.,\cite{Swingle:2014uza}. Non-holographic proposals connecting geometric structures directly to QFT degrees of freedom~\cite{Verlinde:2010hp} aim to broaden this perspective beyond fixed asymptotics.

Stress–tensor deformations provide a concrete probe of such connections. In classical gravity, $T_{\mu\nu}$ sources curvature, whereas in QFT, deformations built from $T_{\mu\nu}$ encode how geometry influences quantum matter. In two dimensions, the solvable $T\bar T$ deformation admits a precise geometric interpretation in terms of random geometry or topological gravity~\cite{Zamolodchikov:2004ce,Smirnov:2016lqw,Cavaglia:2016oda,Dubovsky:2017cnj,Tolley:2019nmm,Cardy:2018sdv}. Higher-dimensional generalizations have been explored~\cite{Taylor:2018xcy,Bonelli:2018kik,Conti:2022egv}, and recent works have constructed geometric realizations of stress–tensor flows without invoking dynamical gravity~\cite{Babaei-Aghbolagh:2024hti,He:2025ppz}. These developments, however, primarily encode classical kinematical geometric features and do not yet incorporate propagating gravitational degrees of freedom.

This Letter formulates a calculable semiclassical correspondence in which the stress-tensor
deformations reorganize a QFT into a gravitational action evaluated at a metric saddle, providing
a geometric representation of stress--tensor deformed QFTs without claiming to derive gravity as a
fundamental emergent phenomenon. Two complementary viewpoints arise. Starting from Einstein or Palatini gravity, one derives the corresponding stress–tensor deformation of the seed QFT. Conversely, starting from deformed theories such as generalized Nambu–Goto and $T\bar T$-like models, one reconstructs the associated gravitational action. 
Beyond the classical saddle, quantum corrections to a nonlocal stress–tensor deformation of massive scalar field theory produces a linearized Einstein gravitational action, showing that the classical metric saddle is the semiclassical limit of the same deformation and ensuring consistency between the classical and quantum levels. 
The correspondence is semiclassical and arises from a single QFT with a nonlocal stress–tensor deformation, distinguishing it from two-sector constructions~\cite{Betzios:2020sro}, metric reformulations~\cite{Conti:2022egv,Morone:2024ffm}, and flows derived from holographic renormalization~\cite{Adami:2025pqr}.

\section{Assumptions and Framework}\label{section General formulas}
The framework relates a stress–tensor–deformed QFT to a corresponding gravitational theory through the path integral, with the deformation controlled by a parameter \( \lambda \). For simplicity, we mainly focus on the following Euclidean gravitational partition function is
\begin{equation}
\mathcal{Z}^{(\lambda)}_{\mathrm{grav}}
= \int \mathcal{D}\psi\,\mathcal{D}g\;
e^{-\hat S[g,\psi] - S^{(\lambda)}_{\mathrm{grav}}[g]},
\label{Basic correspondence}
\end{equation}
where \( \hat S[g,\psi] \) is the seed theory action and \( \psi \) denotes matter fields. The analysis presented in the first part of this Letter is conducted within a semiclassical framework. Evaluating the metric path integral at the saddle gives
\begin{align}
\mathcal{Z}^{(\lambda)}_{\mathrm{grav}}
&=
\int \mathcal{D}\psi
\sum_{g^*}
\mathcal{Z}^{(\lambda)}_{\mathrm{loops}}[g^*]\,
e^{-\hat S[g^*,\psi] - S^{(\lambda)}_{\mathrm{grav}}[g^*]}
\notag\\[-2pt]
&=
\int \mathcal{D}\psi
\sum_{\hat\gamma}\!
\sum_{\alpha}
\mathcal{Z}^{(\lambda)}_{\mathrm{loops}}[g^*_{\alpha}]\,
e^{-S^{(\lambda)}_{\alpha}[\hat\gamma,\psi]},
\label{conjecture 1}
\end{align}
where \(g^*_{\alpha}\to\hat\gamma\) as \(\lambda\to0\), and \(\mathcal{Z}^{(\lambda)}_{\mathrm{loops}}[g^*]\) denotes the fluctuation (loop) contribution around each saddle \( g_{\alpha}^*(\lambda) \) with $g_{\alpha}^*(\lambda)$ and $\hat{\gamma}$ 
satisfying
\begin{align}   
\frac{\delta}{\delta g_{\mu\nu}}
\big[ \hat S[g,\psi] + S^{(\lambda)}_{\mathrm{grav}}[g] \big]
\big|_{g=g^*_{\alpha}} &= 0 \nonumber\\
\frac{\delta}{\delta g_{\mu\nu}}\big[ \hat S[g,\psi] + S^{(\lambda=0)}_{\mathrm{grav}}[g] \big]
\big|_{g=\hat{\gamma}} &= 0.
\label{general gravitational EOM 1}
\end{align}
In this letter, the interpretation (\ref{conjecture 1}) is examined at the classical level. The classical deformed field theory action is related to the gravitational action via 
\begin{equation}
S_{\alpha}^{(\lambda)}[\hat\gamma,\psi]
= \hat S[g^*_{\alpha},\psi] + S^{(\lambda)}_{\mathrm{grav}}[g^*_{\alpha}],
\label{Basic correspondence 1}
\end{equation}

This framework incorporates the random geometric interpretation of two-dimensional $T\bar T$ deformation \cite{Cardy:2018sdv, Tolley:2019nmm}. The corresponding gravitational action is $S^{(\delta\lambda)}_{\mathrm{grav}}=(1/8\delta\lambda)\int d^2x\sqrt{\hat\gamma}\epsilon^{\mu\rho}\epsilon^{\nu\sigma}(g_{\mu\nu}-\hat\gamma_{\mu\nu})(g_{\rho\sigma}-\hat\gamma_{\rho\sigma})$, where $\epsilon^{\mu\nu}$ is the Levi-Civita symbol, and $\hat\gamma$ will eventually emerge as the background metric. By solving the equations of motion (EOM) (\ref{general gravitational EOM 1}), the unique saddle is given by $g^*_{\mu\nu}=\hat\gamma_{\mu\nu}+2\delta\lambda\epsilon_{\mu\rho}\epsilon_{\nu\sigma}\hat T^{\rho\sigma}$, where $\hat T^{\rho\sigma}=\frac{-2}{\sqrt{\hat\gamma}}\frac{\delta \hat S}{\delta\hat\gamma_{\rho\sigma}}$. Substituting it into (\ref{Basic correspondence 1}), the first-order $T\bar T$ deformed action is obtained. In this letter, the aforementioned framework is employed to investigate more general gravitational actions that involve the Ricci curvature and feature a dynamical metric. The construction provides a geometric representation of stress–tensor deformed QFTs, reproducing classical gravitational dynamics at the saddle.


\subsection{Effective field theory action on a deformed metric} 

Within the framework above, once the seed theory $\hat{S}$ and gravitational action $S^{(\lambda)}_{\mathrm{grav}}$ are specified, the deformed QFT action $S^{(\lambda)}_{\alpha}$ can, in principle, be obtained perturbatively from (\ref{Basic correspondence 1}). However, determining the metric saddle point $g^{*}_{\alpha}(\lambda)$ nonperturbatively is generally intractable. Thus, it is useful to define an effective deformed action (EDA) $S_{\mathrm{EDA}}^{(\lambda)}$ on a \emph{deformed metric}, which incorporates the saddle $g_{\alpha}^{*}$. The deformed metric is taken as a local function of $\lambda$ and $g^{*}_{\alpha}$ (more generally, a functional of $g^{*}_{\alpha}$).~\footnote{Although this rewriting amounts to a saddle-level change of variables, it is operationally nontrivial: it recasts the intrinsically bilocal deformation into a covariant derivative expansion on the deformed geometry, yielding (i) a systematic perturbative control of subleading terms and (ii) a unified language for comparison with local
stress--tensor flows of $T\bar T$-type. Once the Green's-function prescription is fixed, scheme-dependent local
counterterms are cleanly separated from the regulator-independent nonlocal kernel.}


A natural choice is to take the metric saddle to be the deformed metric, in which case
\begin{align}
S_{\mathrm{EDA}}^{(\lambda)}[g_{\alpha}^{*},\psi]
&=S_{\alpha}^{(\lambda)}[\hat\gamma,\psi]\notag\\
&=\hat S[g_{\alpha}^{*},\psi]+\int d^dx\mathcal{L}^{(\lambda)}_{\text{st}}(g_{\alpha}^{*\mu\nu},\hat T^*_{\mu\nu}),\label{effective deformed action definition}
\end{align}
where $\mathcal{L}^{(\lambda)}_{\text{st}}$ is a local function of the seed theory stress-tensor obtained by substituting the gravitational EOM into $S_{\mathrm{grav}}^{(\lambda)}$ and eliminating terms associated with curvature tensors. Taking the total derivative of (\ref{Basic correspondence 1}) with respect to \( \lambda \) and using Eq.~(\ref{general gravitational EOM 1}) yields
\begin{equation}
\frac{\text{d}}{\text{d}\lambda}S_{\mathrm{EDA}}^{(\lambda)}[g_{\alpha}^{*},\psi]\equiv\frac{\text{d}}{\text{d}\lambda} S_{\alpha}^{(\lambda)}[\hat\gamma,\psi]
= \partial_{\lambda} S^{(\lambda)}_{\mathrm{grav}}[g^*_{\alpha}] .
\label{flow equation of effective action}
\end{equation}
This formulation provides a direct correspondence between a deformed QFT on the background $\hat{\gamma}$ and a classical gravitational theory. The deformation parameter \( \lambda \) appears solely in the gravitational action, and the flow equation~(\ref{flow equation of effective action}) relates the action's variation to the underlying gravitational dynamics. 


\section{Linearized Einstein gravity as stress tensor deformed theory}\label{subsection perturbation method} 
As a preliminary application, but not limited to, we apply the framework to linearized Einstein gravity. The metric saddle and the induced non-local stress-tensor deformation are derived explicitly. The starting point is the seed theory $\hat S[g,\psi]$ coupled to the Einstein--Hilbert action,
\begin{align}
S=\hat S[g,\psi]+\frac{l^{2}}{2\lambda}\int d^dx\sqrt{g}\,R.
\label{EH action 1}
\end{align}
The saddle-point equation for the metric yields the Einstein field equations with the field theory stress-tensor as source,
\begin{align}
R^*_{\mu\nu}-\frac{1}{2}R^*g^{*}_{\mu\nu}
=-\lambda l^{-2}\hat T_{\mu\nu}^*,
\label{Einstein's equation}
\end{align}
where the subscript $\alpha$ of the metric saddle has been omitted for notational simplicity.

Consider the expansion \( g=\hat\gamma+\lambda h \). To keep the saddle-point fluctuation $h^\ast$ of order $O(\lambda^0)$ and preserve the validity of the perturbative organization, we take the reference background $\hat\gamma_{\mu\nu}$ to satisfy the vacuum Einstein equations, as spelled out in SM A (A5). By employing the standard second-order expansion of the Einstein--Hilbert action~\cite{Gullu:2010em, Altas:2019qcv}, and including the gauge-fixing and ghost terms to handle diffeomorphism redundancy with following standard procedure~\cite{Christensen:1979iy, Giombi:2008vd} for Einstein gravities without cosmological constant~\footnote{For dS and AdS gravities, the gauge-fixing can be found in the recent literature~\cite{Moga:2025sqh}.}, the solution for saddle point in terms of the graviton Green’s function is obtained~\footnote{Solution of metric perturbation $h^*$ follows from a perturbative expansion around a vacuum Einstein background,
which ensures the power counting $h^{\ast}_{\mu\nu}=O(\lambda^0)$ so that higher-order metric terms do not
affect the leading deformation. Total-derivative terms in $h$ are omitted, as they can be absorbed
into boundary terms. Full derivations are provided in SM A (see in particular
Eq.~(A5)). $G_{\mu\nu\rho\sigma}$ is defined as the inverse of the gauge-fixed quadratic graviton operator obtained by expanding the gravitational action to second order around $\hat{\gamma}_{\mu\nu}$. Ghosts contribute only to the one-loop determinant and do not affect the tree-level saddle equation; see SM A.}, $h^*_{\mu\nu}=-2l^{-2}\int d^dy\sqrt{\hat\gamma}G_{\mu\nu\rho\sigma}(\hat T^{\rho\sigma}-\frac{1}{d-2}\hat T\hat\gamma^{\rho\sigma})+O(\lambda)$, with \(G\) satisfying the equation $[-\hat\gamma^{\mu\rho}\hat\gamma^{\nu\sigma}\hat \Box-\hat R^{\mu\rho\nu\sigma}-\hat R^{\nu\rho\mu\sigma}]_xG_{\rho\sigma\alpha\beta}(x,y)=\delta^{(d)}(x\!-\!y)\delta^{\mu}_{\alpha}\delta^{\nu}_{\beta}/\!\sqrt{\!\hat\gamma}$~\footnote{For generic curved background, one can define the reference background $\hat{\gamma}$ satisfies (\ref{general gravitational EOM 1}) and the Green's function can be defined in a classical way which highly depends on the boundary conditions. For simplicity, here we consider the Green’s function on a manifold without boundary, which exhibits exponential decay as the separation between the two points tends to infinity.}.
Substituting the saddle into the action yields the leading deformation of the QFT,
\begin{align}\label{nonlocaldeformation}
S^{(\lambda)}&=\hat S+\frac{\lambda }{2l^2}\!\int\! d^dxd^dy\,\sqrt{\hat\gamma(x)\hat\gamma(y)}G_{\mu\nu\rho\sigma}(x,y)\notag\\
&\quad\times \,
\hat T^{\mu\nu}(x)\Big[\hat T^{\rho\sigma}
-\frac{1}{d-2}\hat T\hat\gamma^{\rho\sigma}\Big](y)+O(\lambda^2).
\end{align}
A standard de Donder gauge-fixing of the metric fluctuation (including the associated ghost term) is employed (SM A (A7)). $G_{\mu\nu\rho\sigma}(x,y)$ is the Green's function of the resulting gauge-fixed spin-2 operator on the reference background $\hat\gamma_{\mu\nu}$, with boundary conditions chosen according to the manifold under consideration.
At the order relevant for (\ref{nonlocaldeformation}), $G_{\mu\nu\rho\sigma}$ enters only through its first part when contracted with the conserved source $\hat T_{\mu\nu}$; gauge-dependent longitudinal components therefore do not affect the deformation.
Similar non-local deformations were studied in \cite{Kawamoto:2025oko}. 
For a flat background \(\hat\gamma_{\mu\nu}=\eta_{\mu\nu}\), the first-order deformation simplifies to
\begin{align}\label{linearnonlocal}
S^{(\lambda)}_{[1]}=\frac{\lambda }{2l^2}\int d^dx\;
\hat T_{\mu\nu}\,\frac{1}{-\hat\Box}\!
\left(\hat T^{\mu\nu}-\frac{1}{d-2}\hat T\eta^{\mu\nu}\right).
\end{align}
The leading deformation organizes as a spin-2 channel exchange controlled by $\hat\square^{-1}$, providing a concrete check against the linearized Einstein response.
The resulting deformation~(\ref{linearnonlocal}) is intrinsically nonlocal on a reference background 
$\hat{\gamma}_{\mu\nu}$ and reproduces the linearized Einstein response for both scalar and Maxwell theories (SM A). 
Higher-order corrections can be computed systematically in perturbation theory (SM A); nonlocality is inevitable at every order in the deformation~\footnote{Since the deformation generates the intrinsically nonlocal kernel~(\ref{linearnonlocal}), the locality assumptions 
entering the Weinberg--Witten theorem do not apply, and the saddle-point correspondence to Einstein dynamics is 
therefore not in conflict with that theorem. Once the Green's function is specified, the operator $1/\Box$ is fixed up to local counterterms; in Minkowski signature, adopting the retarded prescription enforces causality.
}. Moreover, for Gauss-Bonnet gravity, the leading-order deformation retains the same functional form as 
(\ref{nonlocaldeformation}), with the reference metric $\hat{\gamma}_{\mu\nu}$ replaced by the corresponding 
Gauss--Bonnet background. The same saddle-point analysis continues to apply; details are presented in SM A.

In the formulation of effective deformed action (\ref{effective deformed action definition}), substituting~(\ref{Einstein's equation}) into (\ref{EH action 1}) yields $\mathcal{L}^{(\lambda)}_{\text{st}}=\frac{\sqrt{g^*}}{d-2}
\hat{T}^*$. The effective stress-tensor is obtained by taking the functional derivative of $S_{\text{EDA}}$ with respect to the metric, $T_{\mathrm{EDA}}^{(\lambda)\mu\nu}
=\hat{T}^{\mu\nu}
-\frac{1}{d-2}\hat{T}g^{\mu\nu}
-\frac{2}{d-2}\frac{\partial\hat{T}}{\partial g_{\mu\nu}}$. For a massless free scalar seed, $\hat{\mathcal{L}}=\frac{\sqrt{g}}{2}\nabla^{\mu}\phi\nabla_{\mu}\phi$,
one has $g_{\mu\nu}\,\frac{\partial\hat{T}}{\partial g_{\mu\nu}}=\hat{T}$. The trace of the seed theory stress-tensor $\hat T$ can be rewritten in terms of $T_{\text{EDA}}$. Combining this with (\ref{flow equation of effective action}) yields the total $\lambda$-derivative of the effective field action, $\frac{\text{d}}{\text{d}\lambda}S_{\mathrm{EDA}}^{(\lambda)}[g^{*},\psi]
=\frac{1}{4\lambda}\int d^{d}x\sqrt{\!g^{*}}
T_{\mathrm{EDA}}^{(\lambda)*}$. In four-dimensional Maxwell theory the traceless stress tensor preserves conformal invariance, and the linearized Einstein–Hilbert deformation produces the nonlocal action (\ref{nonlocaldeformation}) whose surviving transverse–traceless spin-2 exchange reproduces the standard graviton-mediated backreaction, demonstrating that the metric saddle responds nonlocally to the seed stress tensor and thereby links Einstein dynamics to stress–tensor flows extendable to broader theories.

\section{Non-minimal Palatini gravity v.s. $T\bar{T}$-like flow}
Having established the framework and its realizations in Einstein gravity, we next address the inverse problem: at the classical (saddle) level, we reconstruct non-minimal, metric-affine gravitational actions directly from specified stress-tensor deformations of a seed QFT and verify the correspondence on representative cases via their flow equations. Our discussion is based on the Palatini formalism, in which the Riemann tensor $\mathcal{R}^{\mu}{}_{\nu\rho\sigma}$ and the Ricci tensor $\mathcal{R}_{\mu\nu}$ are constructed with the independent connection. Starting from a specific effective deformed action defined as $S^{(\lambda)}_{\text{EDA}}[g,\psi]=\int d^dx\sqrt{g}\mathcal{B}^{(\lambda)}(g^{\mu\nu}, X_{\mu\nu},\psi)$, where the tensor $X_{\mu\nu}=\sum_{i,j}\mathcal{G}_{ij}(\psi)\nabla_{\mu}\phi^{(i)}\nabla_{\nu}\phi^{(j)}$ and $\mathcal{G}_{ij}$ is the target space metric.
There are $d$ independent invariants constructed from $X_{\mu\nu}$, expressed as $X_n=X^{\mu_n}_{\mu_1}X^{\mu_1}_{\mu_2}\cdots X^{\mu_{n-1}}_{\mu_n}=\text{tr}(X^n)$ for $n=1,2,...,d$. The Lagrangian $\mathcal{B}^{(\lambda)}$ is a local function of $X_n$ and $\phi^{(i)}$, and the effective stress-tensor is given by $(T^{(\lambda)}_{\text{EDA}})^{\mu}_{\nu}=2\sum_{n=1}^{d}n(X^n)^{\mu}_{\nu}\partial_{X_n}\mathcal{B}^{(\lambda)}-\mathcal{B}^{(\lambda)}\delta^{\mu}_{\nu}$.\par
To derive the gravitational theory corresponding to $S^{(\lambda)}_{\text{EDA}}$, we postulate a candidate gravitational action, which is then verified through its equations of motion to check if it reproduces $\mathcal{B}^{(\lambda)}$. Here, we consider a general class of actions with non-minimal couplings between the metric, curvature, and matter fields. The gravitational action~\footnote{Here we assume that the action does not include the Riemann curvature tensor $\mathcal{R^{\mu}_{\nu\rho\sigma}}$. In principle, $\mathcal{A}^{(\mu)}$ should be a local function of all independent invariants composed of $g^{\mu\nu}$, $\mathcal{R}_{\mu\nu}$, and $X^{(i)}_{\mu\nu}$.} can be written as:
\begin{align}
    S^{(\lambda)}_{\text{grav}}[g,\psi]&=\int d^dx\sqrt{g}\mathcal{A}^{(\lambda)}(g^{\mu\nu},\mathcal{R}_{\mu\nu},X_{\mu\nu},\psi), \label{Non-mininal general gravitational action}
\end{align}
where $\mathcal{R}_{\mu\nu}$ is the Ricci tensor in the Palatini formalism. Variations of (\ref{Non-mininal general gravitational action}) with respect to the metric and the independent connection yield
\begin{subequations}
\begin{align}
        \frac{\partial\mathcal{A}^{(\lambda)}}{\partial g^{\mu\nu}}-\frac{1}{2}\mathcal{A}^{(\lambda)}g_{\mu\nu}&=0,\label{non-minimal EoM 1}\\
\bar\nabla_{\sigma}\Big(\sqrt{g}\frac{\partial\mathcal{A}^{(\lambda)}}{\partial\mathcal{R}_{\mu\nu}}\Big)&=0.
\end{align}
\end{subequations}
A suitable gravitational Lagrangian $\mathcal{A}^{(\lambda)}$ yields the deformed field theory Lagrangian $\mathcal{B}^{(\lambda)}$ when evaluated at the metric saddle,
\begin{align}
    S_{\text{grav}}^{(\lambda)}[g^{*},\psi]&=\int d^dx\sqrt{g^{*}}\mathcal{B}^{(\lambda)}(g^{*\mu\nu},X_{\mu\nu},\psi). \label{non-minimal Seff}
\end{align}
The flow equation of the deformed field theory is given by $\partial_{\lambda}S^{(\lambda)}_{\text{EDA}}[g^*,\psi]=\int d^dx\sqrt{g^{*}}\partial_{\lambda}\mathcal{B}^{(\lambda)}(g^{*\mu\nu},X_{\mu\nu},\psi)$.
Different choices of $\mathcal{A}^{(\lambda)}$ lead to different metric saddles $g^{*}$ and hence distinct deformed actions on the reference background. Conversely, given the deformed field theory action on a reference background, the corresponding gravitational action can be reconstructed order by order in metric perturbations. As a realization, we show in SM C that the Generalized Nambu-Goto action arises from a specific Palatini Lagrangian with a non-minimal coupling.

\subsection{Example: $T\bar T$-like deformation in $d$ dimensions}
As a significant example, the $d$-dimensional (root-)$T\bar T$-like deformation introduced in \cite{Babaei-Aghbolagh:2024hti} is investigated, and the corresponding gravitational action is constructed. Suppose that $X^{\mu}_{\nu}$ is diagonalizable, expressed as $X=U\text{diag}(\chi_1,\chi_2,...,\chi_d)U^{-1}$. The effective deformed Lagrangian is presented as follows:
\begin{align}
    \mathcal{B}^{(\lambda)}=\mathcal{B}_0+\lambda^{1-\Sigma}l^{\Delta}\prod_{j=1}^{d}(\chi_j^{\frac{p_j}{2}}-\beta_j^{\frac{p_j}{2}})^{\frac{1}{p_j}},\label{non-minimal effective deformed action}
\end{align}
where $\lbrace{\beta_j}\rbrace$ is the deformation parameters of the root-$T\bar T$-like operator as discussed in \cite{Babaei-Aghbolagh:2024hti}, $\lbrace{p_j}\rbrace$ are the numbers that characterizing the deformation, $\Sigma=\sum_{j=1}^{d}1/p_j$, and $\Delta=(2\Sigma+d-4)d/2$. $\mathcal{B}_0$ satisfies the differential equation $2\chi_j\partial_{\chi_j}\mathcal{B}_0-\mathcal{B}_0=0$ and the solution is given by~\footnote{The solution takes the form
\begin{align}
    \mathcal{B}_0=C(\phi)\prod_{j=1}^{d}\chi_j^{\frac{1}{2}}=C(\phi)\frac{\sqrt{\det(X_{\mu\nu})}}{\sqrt{\det(g_{\mu\nu})}}, \label{NG action}
\end{align}
which is analogous to the action of the Nambu-Goto string \cite{Polchinski:1998rq} up to a regular function $C(\phi)$.}. The effective stress-tensor can be diagonalized by the same matrix $U$, and its eigenvalues $\lbrace{\tau_j}\rbrace$ are given by $\tau_j^{(\lambda)}=2\chi_j\partial_{\chi_j}\mathcal{B}^{(\lambda)}-\mathcal{B}^{(\lambda)}$. These eigenvalues can then be substituted into (\ref{non-minimal effective deformed action}) to derive the flow equation
\begin{align}
    \partial_{\lambda}S_{\text{EDA}}^{(\lambda)}&\!=\!\frac{(1-\Sigma)}{(l^{\Delta}b^{\frac{1}{2}})^{\frac{1}{\Sigma-1}}}\!\int\! d^dx\sqrt{g}\Big(\!\prod_{j=1}^{d}(\tau_j^{(\lambda)})^{\frac{1}{p_j}}\!\Big)^{\frac{1}{\Sigma-1}},
\end{align}
where $b=\prod_{j=1}^{d}\beta_j$. In particular, setting $p_1=p_2=\dots=p_d=p$ reduces the deformation operator to
$\mathcal{O}_{\text{st}}=(\det[(T^{(\lambda)}_{\text{EDA}})^{\mu}_{\nu}])^{\frac{1}{d-p}}$ \cite{Bonelli:2018kik, Cardy:2018sdv}. 

Next, a class of plausible gravitational actions is postulated that could reproduce the effective deformed action given in (\ref{non-minimal effective deformed action}). 
The explicit form of the gravitational Lagrangian $\mathcal{A}^{(\lambda)}$ is introduced as follows:
\begin{align}
    \mathcal{A}^{(\lambda)}&=\mathcal{B}_0+\lambda^{1-\Sigma}l^{\Delta}\prod_{j=1}^{d}\Big(\chi_j^{\frac{p_j}{2}}\notag\\
    \quad&-(\mathfrak{p}_{j}\chi_{j}^{\mathfrak{q}_j}+\mathfrak{s}_jr_{j}^{\mathfrak{q}_j}+\frac{\mathfrak{q}_j-1}{\mathfrak{q}_j}\beta_j)^{\frac{p_j}{2}}\Big)^{\frac{1}{p_j}},\label{non-minimal gravitational action}
\end{align}
where $\lbrace r_j \rbrace$ are the eigenvalues of the Ricci tensor $\mathcal{R}^{\mu}_{\nu}$, assumed to be diagonalizable. $\mathfrak{p}_{j}$, $\mathfrak{q}_{j}$, and $\mathfrak{s}_{j}$ are arbitrary functions independent of $\chi$ and $r$. The EOMs for $\lbrace r_j \rbrace$ follow from Eq.~(\ref{non-minimal EoM 1}); a full derivation is provided in SM D.
Substituting them into (\ref{non-minimal gravitational action}) recovers the effective deformed action in (\ref{non-minimal effective deformed action}). As a significant example, the two-dimensional case of this formula is explicitly analyzed in SM D, which corresponds to the $T\bar T$ flow \cite{Zamolodchikov:2004ce,Smirnov:2016lqw,Cavaglia:2016oda,He:2025ppz}.

\section{Nonlocal Stress--Tensor Deformations and Classical--Quantum Consistency}

Beyond the classical saddle analysis, we examine the nonlocal stress--tensor deformation that underlies our framework and demonstrate its consistency at both classical and quantum levels. 
At the classical level, eliminating the metric perturbation $h_{\mu\nu}$ from the quadratic Einstein--Hilbert action yields a universal bilocal deformation,
\begin{align}
&\delta_{\lambda} S
\!=\!\frac{\lambda}{2l^2}\!\int\!d^dxd^dy\sqrt{\!g(x)g(y)}T^{\mu\nu}(x)H_{\mu\nu\rho\sigma}(x,y)T^{\rho\sigma}(y),\label{eq:classical_kernel}
\end{align}
where the kernel $H_{\mu\nu\rho\sigma}(x,y)=G_{\mu\nu\rho\sigma}(x,y)-1/(d-2)G_{\mu\nu\alpha\beta}(x,y)g^{\alpha\beta}(y)g_{\rho\sigma}(y)$ and the Green's function satisfies $[-g^{\mu\rho}g^{\nu\sigma}\Box-R^{\mu\rho\nu\sigma}-R^{\nu\rho\mu\sigma}]_xG_{\rho\sigma\alpha\beta}(x,y)=\delta(x-y)\delta^{\mu}_{\alpha}\delta^{\nu}_{\beta}$.
Setting $g_{\mu\nu}\to\hat\gamma_{\mu\nu}$ reproduces the leading nonlocal deformation \eqref{nonlocaldeformation} in linearized Einstein gravity.

To probe whether this semiclassical reorganization is compatible with quantum, we compute the leading one-loop
contribution generated by the bilocal interaction~(\ref{eq:classical_kernel}) using the heat-kernel expansion (SM E)~\cite{Seeley:1967ea,DeWitt:1964mxt,Barvinsky:1985an}.
Diffeomorphism covariance forces the induced local effective action into a derivative expansion, 
$\delta\ln \mathcal{Z}^{(\lambda)}[g]\propto\int\sqrt{g}\,(\alpha_{0}+\alpha_{1}R+O(\nabla^{4}))$,
where $\mathcal{Z}^{(\lambda)}[g]$ is the first-order deformed partition function, and the operator content is universal renormalization invariant, but the coefficients $\alpha_{0,1}$ depend on the UV regulator and renormalization scheme through local counterterms. We therefore interpret $\alpha_0$ and $\alpha_1$ as the renormalized vacuum energy and Newton coupling at a chosen reference scale, fixed by a concrete renormalization prescription (SM E).~\footnote{The coefficients of local terms such as $\int\sqrt{g}$ and $\int\sqrt{g}R$ are scheme-dependent and can be
shifted by local counterterms; the nonlocal kernel is fixed once the Green's function prescription is specified.}
. Implementing this prescription in the proper time cutoff scheme yields 
\begin{equation}
\mathcal{Z}^{(\lambda)}_{\mathrm{scalar}}
=\mathcal{Z}_0\Big(1
-\frac{1}{16\pi G_{\text{eff}}}
\int d^4x \sqrt{g}\,
\bigl[
R+O(\nabla^4)
\bigr]\Big),
\label{quantumaction}
\end{equation}
corresponding to the Einstein--Hilbert action supplemented by higher--derivative curvature terms. {The effective Newton's constant is given by $G_{\text{eff}}=\frac{6\pi^3l^2}{\lambda m^4}(\ln(\mu/m))^{-2}$.
The coefficients of local invariants (e.g.\ $\int\!\sqrt{g}$, $\int\!\sqrt{g}\, R$) are renormalization-scheme dependent and can be shifted by local counterterms.}
Consistency of the small--$\lambda$ expansion requires $g|_{\lambda=0}=\hat\gamma$, where $\hat\gamma$ solves the Einstein equations, ensuring that the leading term reproduces the linearized gravitational action \eqref{EH action 1}.

\paragraph*{Classical--quantum consistency.}
Crucially, the classical bilocal kernel \eqref{eq:classical_kernel} and the quantum effective action \eqref{quantumaction} arise from the \emph{same} nonlocal stress--tensor deformation. 
Classically, eliminating the metric yields a bilocal interaction governing linearized gravitational response; quantum mechanically, loop corrections to this interaction generate a local gravitational action with the correct tensorial structure. 
The metric saddle appearing in the classical analysis, therefore, corresponds to the semiclassical (tree-level) limit of the quantum deformation. This demonstrates a coherent, logically self-consistent connection between classical geometric reorganization and quantum gravitational dynamics within a single framework of stress tensor deformation.



\section{Conclusion}

We have formulated a semiclassical correspondence in which stress–tensor deformations of a QFT reorganize into a gravitational action evaluated at the metric saddle. In particular, our results comprise: (i) a universal bilocal kernel equal to the inverse graviton operator reproducing linearized Einstein response; (ii) an explicit inverse construction yielding
Palatini actions from specified stress--tensor deformations; and (iii) a one-loop local curvature term
generated by the same deformation, linking the classical saddle to its quantum completion. The deformation is intrinsically nonlocal, governed by the inverse graviton operator and encoding massless spin-2 exchange.


This nonlocal structure lies outside the locality assumptions of the Weinberg–Witten theorem~\cite{Weinberg:1980kq} and remains compatible with linearized Einstein dynamics. The nonlocal terms also generate quantum corrections to the effective action, offering a controlled setting in which aspects of quantum emergent gravity arise from stress–tensor deformations. 
In Lorentzian applications, the inverse operator in the kernel should be understood with a causal (linear-response) prescription, while a fully nonperturbative assessment of unitarity for the resulting nonlocal deformation lies beyond the scope of the present Letter and will be addressed elsewhere.

Several directions follow naturally.  
(i) \emph{Quantum gravity and holography:} The induced metric flows parallel holographic renormalization~\cite{deBoer:1999tgo,Skenderis:2002wp}, suggesting routes toward emergent de Sitter or flat-space holography~\cite{Strominger:2001pn,Anninos:2011ui} and offering a complementary perspective on information-theoretic aspects of gravity~\cite{Ryu:2006bv,Hubeny:2007xt}.  
(ii) \emph{Spacetime gluing:} The consistent gluing of stress–tensor deformed QFTs suggests an analogue for gluing spacetimes~\cite{Shen:2024itl,Kawamoto:2023wzj,Kawamoto:2025oko}, where implementing junction conditions is typically nontrivial.

\begin{acknowledgments}
We thank Ronggen Cai, Mingzhe Li,  Roberto Tateo, Stefan Theisen, Tianhao Wu, and Lixin Xu for helpful discussions. We are particularly grateful to Tianhao Wu for informing us to apply the heat-kernel to enable our completely independent implementation of the main-text derivation, whereby a massive scalar field, under a deformation, yields linearized Einstein gravity. This work was supported by NSFC Grant Nos. 12475053, 12588101, and 12235016.\par


\end{acknowledgments}

\bibliographystyle{apsrev4-2}
\bibliography{scalartoLiouville}


\newpage

\onecolumngrid
 
\appendix
\setcounter{section}{1}
\setcounter{equation}{0}
\begin{center}{\large \textbf{SUPPLEMENTAL MATERIAL
\\\vspace{0.1cm}
}
}
\end{center}

 \appendix
\setcounter{section}{1}
\setcounter{equation}{0}
 %

\begin{center}{\large 
\textbf{A.  Exact deformed action from perturbation method }}
\end{center}
\vspace{0.15cm}

To determine the exact form of the deformed theory, we compute the metric saddle point $g^{*}$ and express the deformed action in terms of the background metric $\hat{\gamma}$. Consider the metric perturbation $g=\hat\gamma+\lambda h$. By employing the techniques in \cite{Gullu:2010em,Altas:2019qcv}, we can expand the Einstein-Hilbert action to the second order of $h$,
\begin{align}
    S_{\text{EH}}&=l^{2}\int d^dx\sqrt{\hat\gamma}\Big[\hat R+\Big(R^{(1)}+\frac{\lambda}{2}\hat R\hat\gamma^{\mu\nu}h_{\mu\nu}\Big)\notag\\
    &\quad+\Big(R^{(2)}+\frac{\lambda}{2}R^{(1)}\hat\gamma^{\mu\nu}h_{\mu\nu}+\frac{\lambda}{8}\hat Rh_{\mu\nu}(\hat\gamma^{\mu\nu}\hat\gamma^{\rho\sigma}-2\hat\gamma^{\mu\rho}\hat\gamma^{\nu\sigma})h_{\rho\sigma}\Big)+O(\lambda^3 h^3)\Big].
\end{align}
The perturbations of the Ricci scalar curvature are
\begin{align}
    R^{(1)}&=\lambda\hat \nabla^{\mu}\hat \nabla^{\nu}h_{\mu\nu}-\lambda\hat \Box(\hat\gamma^{\mu\nu}h_{\mu\nu})-\lambda\hat R^{\mu\nu}h_{\mu\nu},\notag\\
    R^{(2)}&=\lambda^2\hat R^{\mu\rho}\hat\gamma^{\nu\sigma}h_{\mu\nu}h_{\rho\sigma}-\frac{\lambda^2}{2}h_{\mu\nu}\hat\gamma^{\nu\sigma}\hat \nabla^{\rho}\hat \nabla^{\mu}h_{\rho\sigma}+\frac{\lambda^2}{4}h_{\mu\nu}\hat\gamma^{\mu\rho}\hat\gamma^{\nu\sigma}\hat \Box h_{\rho\sigma}\notag\\
    &\quad+\frac{\lambda^2}{4}h_{\mu\nu}\hat\gamma^{\mu\nu}\hat\gamma^{\rho\sigma}\hat \Box h_{\rho\sigma}+(\text{total derivatives}).
\end{align}
It follows that
\begin{align}\label{SM A3}
    S_{\text{EH}}&=l^{2}\int d^dx\sqrt{\hat\gamma}\Big[\hat R+\lambda\Big(\frac{1}{2}\hat R\hat\gamma^{\mu\nu}h_{\mu\nu}-\hat R^{\mu\nu}h_{\mu\nu}\Big)\notag\\
    &\quad+\Big(\frac{\lambda^2}{4}h_{\mu\nu}(\hat\gamma^{\mu\rho}\hat\gamma^{\nu\sigma}-\hat\gamma^{\mu\nu}\hat\gamma^{\rho\sigma})\hat \Box h_{\rho\sigma}+\frac{\lambda^2}{2}h_{\mu\nu}\hat\gamma^{\rho\sigma}\hat \nabla^{\mu}\hat \nabla^{\nu}h_{\rho\sigma}-\frac{\lambda^2}{2}h_{\mu\nu}\hat\gamma^{\mu\rho}\hat \nabla^{\sigma}\hat \nabla^{\nu}h_{\rho\sigma}\notag\\
    &\quad+\lambda^2h_{\mu\nu}[\hat R^{\mu\rho}\hat\gamma^{\nu\sigma}-\frac{1}{2}\hat R^{\mu\nu}\hat\gamma^{\rho\sigma}-\frac{1}{8}\hat R(2\hat\gamma^{\mu\rho}\hat\gamma^{\nu\sigma}-\hat\gamma^{\mu\nu}\hat\gamma^{\rho\sigma})]h_{\rho\sigma}\Big)+O(\lambda^3h^3)\Big].
\end{align}
Taking variation of this expansion with respect to $h_{\mu\nu}$, we find the EOM for auxiliary field, 
\begin{align}
    &\lambda\Big[\frac{1}{2}(\hat\gamma^{\mu\rho}\hat\gamma^{\nu\sigma}-\hat\gamma^{\mu\nu}\hat\gamma^{\rho\sigma})\hat \Box+\frac{1}{2}(\hat\gamma^{\rho\sigma}\hat \nabla^{\mu}\hat \nabla^{\nu}+\hat\gamma^{\mu\nu}\hat \nabla^{\rho}\hat \nabla^{\sigma})-\frac{1}{2}\hat\gamma^{\mu\rho}\hat \nabla^{\sigma}\hat \nabla^{\nu}-\frac{1}{2}\hat\gamma^{\nu\rho}\hat \nabla^{\sigma}\hat \nabla^{\mu}\Big]h_{\rho\sigma}\notag\\
    &+\lambda\Big[\hat R^{\mu\rho}\hat\gamma^{\nu\sigma}+\hat R^{\nu\rho}\hat\gamma^{\mu\sigma}-\frac{1}{2}\hat R^{\mu\nu}\hat\gamma^{\rho\sigma}-\frac{1}{2}\hat R^{\rho\sigma}\hat\gamma^{\mu\nu}-\frac{1}{4}\hat R(2\hat\gamma^{\mu\rho}\hat\gamma^{\nu\sigma}-\hat\gamma^{\mu\nu}\hat\gamma^{\rho\sigma})\Big]h_{\rho\sigma}\notag\\
    &+\frac{1}{2}\hat R\hat\gamma^{\mu\nu}-\hat R^{\mu\nu}-l^{-2}\lambda \hat T^{\mu\nu}- l^{-2}\lambda ^2h_{\rho\sigma}M^{\mu\nu\rho\sigma}=0,
\end{align}
where $M^{\mu\nu\rho\sigma}=\frac{\partial \hat T^{\mu\nu}}{\partial\hat\gamma_{\rho\sigma}}+\frac{\partial \hat T^{\rho\sigma}}{\partial\hat\gamma_{\mu\nu}}+\frac{1}{2}\hat T^{\mu\nu}\hat\gamma^{\rho\sigma}+\frac{1}{2}\hat T^{\rho\sigma}\hat\gamma^{\mu\nu}$. To ensure that the higher-order terms of the metric perturbation do not affect the leading-order contribution to the stress tensor deformation and to maintain the validity of the perturbation analysis, the saddle point $h^*_{\mu\nu}$ should be proportional to $O(\lambda^0)$ (which will be explained in detail in the next subsection). This condition requires that the background metric satisfy the vacuum Einstein field equations,
\begin{align}
    \hat R^{\mu\nu}-\frac{1}{2}\hat R\hat\gamma^{\mu\nu}=0. \label{vacuum Einstein equation}
\end{align}
By introducing the trace-reversed variable
\begin{align}
    \tilde{h}_{\rho\sigma}=h_{\rho\sigma}-\frac{1}{2}\hat\gamma_{\rho\sigma}h^{\alpha}_{\alpha},
\end{align}
and incorporating the following gauge fixing term and ghost term into the gravitational action \cite{Christensen:1979iy,Giombi:2008vd},
\begin{align}\label{Gauge fixing and ghost}
    S_{\text{gauge}}&=-\frac{\lambda^2l^{2}}{2}\int d^dx\sqrt{\hat\gamma}\hat \gamma^{\nu\rho}\hat \nabla^{\mu}\tilde h_{\mu\nu}\hat \nabla^{\sigma}\tilde h_{\rho\sigma},\notag\\
    S_{\text{ghost}}&=-\frac{l^{2}}{2}\int d^dx\sqrt{\hat\gamma}\bar \eta_{\mu}(\hat\gamma^{\mu\nu}\hat \Box+\hat R^{\mu\nu})\eta_{\nu},
\end{align}
the above EOM can be simplified as
\begin{align}
    \frac{1}{2}\hat \Box h^{\mu\nu}-\frac{1}{4}\hat\gamma^{\mu\nu}\hat \Box h^{\alpha}_{\alpha}+\frac{1}{2}(\hat R^{\mu\rho\nu\sigma}+\hat R^{\nu\rho\mu\sigma})h_{\rho\sigma}-l^{-2}\hat T^{\mu\nu}-l^{-2}\lambda h_{\rho\sigma}M^{\mu\nu\rho\sigma}=0.
\end{align}
The corresponding solution for auxiliary field takes the form 
\begin{align}
    h^*_{\mu\nu}(x)&=-\frac{2}{l^2} \int d^dy\sqrt{\hat\gamma(y)}G_{\mu\nu\rho\sigma}(x,y)\Big(\hat T^{\rho\sigma}(y)-\frac{1}{d-2}\hat T(y)\hat\gamma^{\rho\sigma}(y)\Big)+O(\lambda),\label{Einstein gravity metric saddle perturbation 1}
\end{align}
Plugging it into the gravitational action, we obtain the leading-order contribution to the deformed action,
\begin{align}
        S^{(\lambda)}[\hat\gamma,\psi]&=\hat S[\hat\gamma,\psi]+\frac{\lambda l^{-2}}{2}\int d^dxd^dy\sqrt{\hat\gamma(x)\hat\gamma(y)}G_{\mu\nu\rho\sigma}(x,y)\hat T^{\mu\nu}(x)[\hat T^{\rho\sigma}-\frac{1}{d-2}\hat T^{\alpha}_{\alpha}\hat\gamma^{\rho\sigma}](y)\notag\\
        &\quad+O(\lambda^2).
\end{align}
A simpler case is that the background metric is flat, $\hat\gamma_{\mu\nu}=\eta_{\mu\nu}$. The Green's function
\begin{align}
    G_{\mu\nu\alpha\beta}(x,y)&=\eta_{\mu\alpha}\eta_{\nu\beta}G_{(-\hat\Box)}(x,y).
\end{align}
Then, the leading order deformation can be simplified as
\begin{align}\label{firstleadingdeformation}
    S^{(\lambda)}_{[1]}[\hat\gamma,\psi]=\frac{\lambda l^{-2}}{2}\int d^dx\sqrt{\hat\gamma}\hat T_{\mu\nu}\frac{1}{-\hat\Box}(\hat T^{\mu\nu}-\frac{1}{d-2}\hat T^{\alpha}_{\alpha}\eta^{\mu\nu}).
\end{align}\par
{ As an illustrated example, we consider the first-order non-local stress tensor deformation of the free scalar field theory when the background metric is flat,
\begin{align}\label{first order deformed action}
    S^{(\lambda)}[\eta,\phi]&=\hat S[\eta,\phi]+\frac{\lambda l^{-2}}{2}\int d^dx\sqrt{\eta}\hat T_{\mu\nu}\frac{1}{-\hat\Box}(\hat T^{\mu\nu}-\frac{1}{d-2}\hat T^{\alpha}_{\alpha}\eta^{\mu\nu})+O(\lambda^2)\notag\\
    &=\frac{1}{2}\int d^dx\sqrt{\eta}\Big[\eta^{\rho\sigma}\partial_{\rho}\phi\partial_{\sigma}\phi+\lambda l^{-2}\big(\partial_{\mu}\phi\partial_{\nu}\phi-\frac{1}{2}\eta_{\mu\nu}\eta^{\rho\sigma}\partial_{\rho}\phi\partial_{\sigma}\phi\big)\frac{1}{-\hat\Box}\partial^{\mu}\phi\partial^{\nu}\phi\Big]+O(\lambda^2).
\end{align}
By introducing the metric redefinition
\begin{align}
    g_{\mu\nu}&=\eta_{\mu\nu}-2\lambda l^{-2}\frac{1}{-\hat\Box}\partial_{\mu}\phi\partial_{\nu}\phi+O(\lambda^2),\notag\\
    g^{\mu\nu}&=\eta^{\mu\nu}+2\lambda l^{-2}\frac{1}{-\hat\Box}\partial^{\mu}\phi\partial^{\nu}\phi+O(\lambda^2),
\end{align}
and plugging them into the first order deformed action (\ref{first order deformed action}), we obtain
\begin{align}
S^{(\lambda)}&=\frac{1}{2}\int d^dx\sqrt{g}g^{\rho\sigma}\partial_{\rho}\phi\partial_{\sigma}\phi-\frac{l^2}{8\lambda}\int d^dx\sqrt{\eta}\Big((g^{\mu\nu}-\eta^{\mu\nu})\hat\Box(g_{\mu\nu}-\eta_{\mu\nu})-\frac{1}{2}(\eta_{\mu\nu}g^{\mu\nu}-d)\hat\Box(\eta^{\rho\sigma}g_{\rho\sigma}-d)\Big)+O(\lambda^2),
\end{align}
which can be regarded as the free scalar field theory on the metric $g$ coupled with the linearized Einstein's gravity.\par
A further illustrative example is that of Einstein-Maxwell theory, whose gravitational action is given by
\begin{align}
    S_{\text{grav}}^{(\lambda)}[g,A]=-\frac{1}{4}\int d^4x\sqrt{-g}F_{\mu\nu}F^{\mu\nu}+\frac{1}{2\lambda}\int d^4x\sqrt{-g}R.
\end{align}
This system is first analyzed using the method of the effective deformed action introduced in the main text. Taking the variation with respect to the metric, we find the Einstein's equation
\begin{align}
    R_{\mu\nu}-\frac{1}{2}Rg_{\mu\nu}=2\lambda \hat T_{\mu\nu},
\end{align}
where
\begin{align}
    \hat T_{\mu\nu}=F_{\mu\alpha}F_{\nu}{}^{\alpha}-\frac{1}{4}g_{\mu\nu}F_{\alpha\beta}F^{\alpha\beta}.
\end{align}
The metric solution is denoted as $g^{(\lambda)}$. Taking the trace of the Einstein's equation, and using the fact that $g^{\mu\nu}\hat T_{\mu\nu}=0$, we find
\begin{align}
    R(g^{(\lambda)})=\hat T(g^{(\lambda)})=0.
\end{align}
Consequently, the effective deformed theory retains the same functional form as Maxwell's theory, while the metric is replaced by the solution to the Einstein's equation.
\begin{align}
    S_{\text{EDA}}[g^{(\lambda)},A]=\Big(S_{\text{Maxwell}}[g,A]+S^{(\lambda)}_{\text{EH}}[g]\Big)\Big|_{g=g^{(\lambda)}}=S_{\text{Maxwell}}[g^{(\lambda)},A].
\end{align}
We can further prove that when the metric satisfies the Einstein's equation, the effective deformed action is independent of the deformation parameter $\lambda$, i.e.,
\begin{align}
    \frac{\text{d}}{\text{d}\lambda}S_{\text{EDA}}[g^{(\lambda)},A]=\frac{\text{d}}{\text{d}\lambda}S_{\text{Maxwell}}[g^{(\lambda)},A]=0.
\end{align}
There are two approaches to prove this identity. The first is to consider the derivative of the gravitational action $S^{(\lambda)}[g^{(\lambda)},A]$ with respect to the deformation parameter,
\begin{align}
    \frac{\text{d}}{\text{d}\lambda}S_{\text{EDA}}[g^{(\lambda)},A]&=\frac{\text{d}}{\text{d}\lambda}\Big(S_{\text{Maxwell}}[g,A]+S^{(\lambda)}_{\text{EH}}[g]\Big)_{g=g^{(\lambda)}}\notag\\
    &=\frac{\partial}{\partial\lambda}S^{(\lambda)}_{\text{EH}}[g^{(\lambda)}]+\int d^4x\frac{\delta(S_{\text{Maxwell}}+S^{(\lambda)}_{\text{EH}})}{\delta g^{\mu\nu}}\Big|_{g=g^{(\lambda)}}\frac{\partial g^{(\lambda)\mu\nu}}{\partial\lambda}\notag\\
    &=-\frac{1}{2\lambda^2}\int d^4x\sqrt{-g^{(\lambda)}}R(g^{(\lambda)})=0.
\end{align}
The second approach examines the action of Maxwell theory on the metric saddle $g^{(\lambda)}$. Since the stress tensor of Maxwell theory is traceless, this yields an integral identity
\begin{align}
    \int d^4x\sqrt{-g^{(\lambda)}}R(g^{(\lambda)})=0.
\end{align}
Consider a small variation of the deformation parameter, $\lambda\to\lambda+\delta\lambda$, we find
\begin{align}
    \delta\int d^4x\sqrt{-g^{(\lambda)}}R(g^{(\lambda)})=\int d^4x\sqrt{-g^{(\lambda)}}\Big(R_{\mu\nu}(g^{(\lambda)})-\frac{1}{2}R(g^{(\lambda)})g^{(\lambda)}_{\mu\nu}\Big)\frac{\partial g^{(\lambda)\mu\nu}}{\partial\lambda}\delta\lambda=0.
\end{align}
Here, we consider an unbounded spacetime and neglect the boundary terms in the variation of the action. From the Einstein's equation, we have
\begin{align}
    \int d^4x\sqrt{-g^{(\lambda)}}\Big(R_{\mu\nu}(g^{(\lambda)})-\frac{1}{2}R(g^{(\lambda)})g^{(\lambda)}_{\mu\nu}\Big)\frac{\partial g^{(\lambda)\mu\nu}}{\partial\lambda}=2\lambda\int d^4x\sqrt{-g^{(\lambda)}}\hat T_{\mu\nu}(g^{(\lambda)})\frac{\partial g^{(\lambda)\mu\nu}}{\partial\lambda}=0.
\end{align}
We now return to the action of Maxwell theory. For nonzero $\lambda$, taking the first order derivative with respect to $\lambda$ yields
\begin{align}
    \frac{\text{d}}{\text{d}\lambda}S_{\text{Maxwell}}[g^{(\lambda)},A]&=\int d^4x\frac{\delta S_{\text{Maxwell}}}{\delta g^{\mu\nu}}\Big|_{g=g^{(\lambda)}}\frac{\partial g^{(\lambda)\mu\nu}}{\partial\lambda}\notag\\
    &=\frac{1}{2}\int d^4x\sqrt{-g^{(\lambda)}}\hat T_{\mu\nu}(g^{(\lambda)})\frac{\partial g^{(\lambda)\mu\nu}}{\partial\lambda}=0.
\end{align}
To demonstrate more explicitly that the Maxwell theory on the deformed metric is independent of the deformation parameter, we consider a static and spherically symmetric metric solution. Given the assumptions of time translation invariance and rotational invariance for the electromagnetic field tensor, it can possess at most two non-vanishing independent components \cite{Wald:1984rg,Carroll:1997ar,Carroll:2004st}. The electromagnetic two-form can be formally written as
\begin{align}
    F=A(r)\ dt\wedge dr+B(r)\sin\theta\ d\theta\wedge d\varphi.
\end{align}
By employing the Maxwell equation, the coefficients of $F$ take the forms
\begin{align}
    A(r)=-\frac{Q_e}{r^2},\ \ \ \ B(r)=Q_m.
\end{align}
The metric solution is assumed to adopt the conventional static and spherically symmetric form,
\begin{align}
    ds^2=-f^{(\lambda)}(r)dt^2+\frac{dr^2}{f^{(\lambda)}(r)}+r^2(d\theta^2+\sin^2\theta\ d\varphi^2).
\end{align}
The non-vanishing components of the inverse metric are $g^{(\lambda)tt}=-\frac{1}{f^{(\lambda)}(r)}$, $g^{(\lambda)rr}=f^{(\lambda)}(r)$, $g^{(\lambda)\theta\theta}=\frac{1}{r^2}$, and $g^{(\lambda)\varphi\varphi}=\frac{1}{r^2\sin^2\theta}$. It follows that
\begin{align}
    F^{tr}=-A(r),\ \ \ F^{\theta\varphi}=\frac{B(r)}{r^4\sin\theta}.
\end{align}
By plugging the components of the electromagnetic field and the metric into Einstein's equation, we obtain the explicit form of $f^{(\lambda)}$, which is exactly the Reissner-Nordstr\"{o}m solution \cite{Carroll:1997ar},
\begin{align}
    f^{(\lambda)}(r)=1-\frac{\lambda M}{4\pi r}+\frac{\lambda(Q_e^2+Q_m^2)}{8\pi r^2}.
\end{align}
Putting everything together, and using the definition of effective deformed field theory action, we find
\begin{align}
    S_{\text{EDA}}[g^{(\lambda)},Q_m,Q_e]=\int dtdrd\theta d\varphi\ \frac{\sin\theta(Q_m^2-Q_e^2)}{r^2},
\end{align}
which is independent of $\lambda$.\par
It is crucial to distinguish the above result carefully from that derived using a perturbative approach. Analogous to the scalar field case, we investigate the first-order non-local stress tensor deformation of Maxwell theory on a flat background,
\begin{align}
    S^{(\lambda)}[\eta,A]&=S_{\text{Maxwell}}[\eta,A]+\frac{\lambda l^{-2}}{2}\int d^4x\sqrt{-\eta}\hat T_{\mu\nu}\frac{1}{-\hat\Box}(\hat T^{\mu\nu}-\frac{1}{2}\hat T^{\alpha}_{\alpha}\eta^{\mu\nu})+O(\lambda^2)\notag\\
    &=\int d^4x\sqrt{-\eta}\Big[-\frac{1}{4}F_{\alpha\beta}F^{\alpha\beta}+\frac{\lambda l^{-2}}{2}(F_{\mu\alpha}F_{\nu}{}^{\alpha}-\frac{1}{4}\eta_{\mu\nu}F_{\alpha\beta}F^{\alpha\beta})\frac{1}{-\hat\Box}(F^{\mu}{}_{\alpha}F^{\nu\alpha}-\frac{1}{4}\eta^{\mu\nu}F_{\alpha\beta}F^{\alpha\beta})\Big]+O(\lambda^2).
\end{align}
By introducing the metric redefinition
\begin{align}
    g_{\mu\nu}&=\eta_{\mu\nu}+2\lambda l^{-2}\frac{1}{-\hat\Box}(F_{\mu\alpha}F_{\nu}{}^{\alpha}-\frac{1}{4}\eta_{\mu\nu}F_{\alpha\beta}F^{\alpha\beta})+O(\lambda^2),\notag\\
    g^{\mu\nu}&=\eta^{\mu\nu}-2\lambda l^{-2}\frac{1}{-\hat\Box}(F^{\mu}{}_{\alpha}F^{\nu\alpha}-\frac{1}{4}\eta^{\mu\nu}F_{\alpha\beta}F^{\alpha\beta})+O(\lambda^2),
\end{align}
the deformed field theory action can be rewritten as
\begin{align}
    S^{(\lambda)}&=-\frac{1}{4}\int d^4x\sqrt{-g}g^{\alpha\gamma}g^{\beta\delta}F_{\alpha\beta}F_{\gamma\delta}-\frac{l^2}{8\lambda}\int d^4x\sqrt{-\eta}(g^{\mu\nu}-\eta^{\mu\nu})\hat\Box(g_{\mu\nu}-\eta_{\mu\nu})+O(\lambda^2).
\end{align}
This deformed action can be interpreted as a coupling between Maxwell theory on the deformed metric and linearized gravity. In contrast to the EFT approach, the linearized gravity sector here encompasses both the linearized Einstein–Hilbert term and the gauge-fixing term introduced in (\ref{Gauge fixing and ghost}), (more precisely, the action should also incorporate additional ghost terms.) As a result, the first-order derivative of the deformed action with respect to $\lambda$ is non-vanishing.}
\par
The Einstein-Hilbert action contains higher-order corrections of $h_{\mu\nu}$. In principle, for the perturbation method to be valid, we must demonstrate that the higher-order corrections of $h_{\mu\nu}$ do not affect the leading-order form of the stress tensor deformation. The gravitational action can be formally written as
\begin{align}
    S&=\hat S[\hat\gamma+\lambda h,\psi]+\frac{1}{2\lambda}S_{\text{grav}}[\hat\gamma+\lambda h]\notag\\
    &=\hat S[\hat\gamma,\psi]-\frac{\lambda }{2}\int d^dx\sqrt{\hat\gamma}h_{\mu\nu}\hat T^{\mu\nu}-\frac{1}{2}\sum_{k=1}^{\infty}\lambda ^{k+1}\int d^dxh_{\mu\nu}\Big(\prod_{i=1}^{k}h_{\mu_i\nu_i}\Big)\frac{\partial^k}{\prod_{i=1}^k\partial\hat\gamma_{\mu_i\nu_i}}\Big(\sqrt{\hat\gamma}\hat T^{\mu\nu}\Big)\notag\\
    &\quad+\frac{l^{2}}{2\lambda}\sum_{k=0}^{\infty}\int d^dx\sqrt{\hat\gamma}\mathcal{F}^{(k)},
\end{align}
where
\begin{align}
    \mathcal{F}^{(k)}&=\mathcal{F}^{(k)}(\hat\gamma_{\mu\nu},h_{\mu\nu},\hat\nabla_{\rho} h_{\mu\nu},\hat\nabla_{\rho}\hat\nabla_{\sigma}h_{\mu\nu})\sim \lambda^kh^k,\notag\\
    [\mathcal{F}^{(k)}]&=[L]^{-2}.
\end{align}
The EOM for the auxiliary field is
\begin{align}
    &-\lambda l^{-2}\Big[{\hat T^{\mu\nu}+\sum_{k=1}^{\infty}\Big(\lambda^k\prod_{i=1}^{k}h_{\mu_i\nu_i}\Big)\frac{1}{\sqrt{\hat\gamma}}\Big(\frac{\partial^k}{\prod_{i=1}^k\partial\hat\gamma_{\mu_i\nu_i}}(\sqrt{\hat\gamma}\hat T^{\mu\nu})+k\frac{\partial^k}{\partial\hat\gamma_{\mu\nu}\prod_{i=2}^k\partial\hat\gamma_{\mu_i\nu_i}}(\sqrt{\hat\gamma}\hat T^{\mu_1\nu_1})\Big)}\Big]\notag\\
    &\quad+\sum_{k=0}^{\infty}\tilde{\mathcal{F}}^{(k)\mu\nu}=0,
\end{align}
where
\begin{align}
    \tilde{\mathcal{F}}^{(k)\mu\nu}(x)=\frac{1}{\sqrt{\hat\gamma(x)}}\frac{\delta}{\lambda\delta h_{\mu\nu}(x)}\int d^dx'\sqrt{\hat\gamma}\mathcal{F}^{(k+1)}\sim \lambda^kh^k.
\end{align}
The first two coefficients have been calculated in the previous section,
\begin{align}
    \tilde{\mathcal{F}}^{(0)\mu\nu}&=\frac{1}{2}\hat R\hat\gamma^{\mu\nu}-\hat R^{\mu\nu},\notag\\
    \tilde{\mathcal{F}}^{(1)\mu\nu}&=\frac{\lambda}{2}\hat \Box h^{\mu\nu}-\frac{\lambda}{4}\hat\gamma^{\mu\nu}\hat \Box h^{\alpha}_{\alpha}+\frac{\lambda}{2}\Big[\hat R^{\mu\rho\nu\sigma}+\hat R^{\nu\rho\mu\sigma}+\hat R^{\mu\rho}\hat\gamma^{\nu\sigma}+\hat R^{\nu\rho}\hat\gamma^{\mu\sigma}\notag\\
    &\quad-\hat R^{\mu\nu}\hat\gamma^{\rho\sigma}-\hat R^{\rho\sigma}\hat\gamma^{\mu\nu}-\frac{1}{2}\hat R(2\hat\gamma^{\mu\rho}\hat\gamma^{\nu\sigma}-\hat\gamma^{\mu\nu}\hat\gamma^{\rho\sigma})\Big]h_{\rho\sigma}.
\end{align}
Suppose that the solution can be written as a power series in $\lambda$,
\begin{align}
    h^*_{\mu\nu}=\sum^{\infty}_{k=-\infty}\lambda^kh^{*}_{[k]\mu\nu}.
\end{align}
It is clear that when $\tilde{\mathcal{F}}^{(0)}_{\mu\nu}=0$, there exists a solution of the auxiliary field that satisfies
\begin{align}
    h^{*}_{[k]\mu\nu}=0,\text{ for }k<0.
\end{align}
Such a saddle point corresponds to the seed theory with a non-local stress tensor deformation, and the form of the deformation can be solved order by order using the perturbation method. Below, we proceed to calculate the deformation up to the second order in $\lambda$. One can easily find that
\begin{align}
\tilde{\mathcal{F}}^{(2)\mu\nu}&=\frac{\lambda^2}{8}\Big(4\hat\nabla_{\rho}h^{\mu\nu}\hat\nabla_{\sigma}h^{\rho\sigma}-8\hat\nabla_{\rho}h^{\mu\rho}\hat\nabla_{\sigma}h^{\nu\sigma}-2h^{\mu\nu}\hat\nabla_{\sigma}\hat\nabla_{\rho}h^{\rho\sigma}+8h^{\nu\rho}\hat\nabla_{\sigma}\hat\nabla_{\rho} h^{\mu\sigma}+4h^{\rho\sigma}\hat\nabla_{\sigma}\hat\nabla_{\rho}h^{\mu\nu}+2h^{\mu\nu}\hat\Box h^{\rho}_{\rho}\notag\\
&\quad-8h^{\nu\rho}\hat\Box h^{\mu}_{\rho}+2h^{\rho}_{\rho}\hat\Box h^{\mu\nu}-8h^{\mu\rho}(\hat\nabla_{\rho}\hat\nabla_{\sigma}h^{\nu\sigma}+\hat\nabla_{\sigma}\hat\nabla_{\rho}h^{\nu\sigma}-\hat\Box h^{\nu}_{\rho}-\hat\nabla_{\sigma}\hat\nabla^{\nu}h_{\rho}^{\sigma})-4h^{\rho}_{\rho}\hat\nabla_{\sigma}\hat\nabla^{\nu}h^{\mu\sigma}-\hat\gamma^{\mu\nu}\hat\nabla_{\sigma}h^{\delta}_{\delta}\hat\nabla^{\sigma}h^{\rho}_{\rho}\notag\\
&\quad+8\hat\nabla_{\sigma}h^{\nu}_{\rho}\hat\nabla^{\sigma}h^{\mu\rho}-4\hat\gamma^{\mu\nu}\hat\nabla_{\rho}h^{\rho\sigma}\hat\nabla_{\delta}h_{\sigma}^{\delta}+4\hat\gamma^{\mu\nu}\hat\nabla^{\sigma}h^{\rho}_{\rho}\hat\nabla_{\delta}h_{\sigma}^{\delta}+2\hat\gamma^{\mu\nu}h^{\rho}_{\rho}\hat\nabla_{\delta}\hat\nabla_{\sigma}h^{\sigma\delta}-2\hat\gamma^{\mu\nu}h^{\rho}_{\rho}\hat\Box h^{\sigma}_{\sigma}-2\hat\gamma^{\mu\nu}\hat\nabla_{\sigma}h_{\rho\delta}\hat\nabla^{\delta}h^{\rho\sigma}\notag\\
&\quad+3\hat\gamma^{\mu\nu}\hat\nabla_{\delta}h_{\rho\sigma}\hat\nabla^{\delta}h^{\rho\sigma}+4\hat\nabla_{\rho}h^{\nu\rho}\hat\nabla^{\mu}h^{\sigma}_{\sigma}+4h^{\rho}_{\rho}\hat\nabla^{\mu}\hat\nabla_{\sigma}h^{\nu\sigma}-4h^{\rho\sigma}\hat\nabla^{\mu}\hat\nabla^{\nu}h_{\rho\sigma}-8\hat\nabla^{\sigma}h^{\mu\rho}\hat\nabla^{\nu}h_{\rho\sigma}+4\hat\nabla^{\mu}h^{\rho\sigma}\hat\nabla^{\nu}h_{\rho\sigma}\notag\\
&\quad+4\hat\nabla_{\rho}h^{\mu\rho}\hat\nabla^{\nu}h^{\sigma}_{\sigma}-2\hat\nabla^{\mu}h^{\rho}_{\rho}\hat\nabla^{\nu}h^{\sigma}_{\sigma}-2\hat\nabla_{\rho}h^{\sigma}_{\sigma}(\hat\nabla^{\rho}h^{\mu\nu}-2\hat\nabla^{\nu}h^{\mu\rho})-8\hat\nabla_{\sigma}h_{\rho}^{\sigma}\hat\nabla^{\nu}h^{\mu\rho}-4h^{\rho}_{\rho}\hat\nabla^{\nu}\hat\nabla_{\sigma}h^{\mu\sigma}+2h^{\rho}_{\rho}\hat\nabla^{\nu}\hat\nabla^{\mu}h^{\sigma}_{\sigma}\Big).
\end{align}
The saddle of the auxiliary field to the second order in $\lambda$ takes the form
\begin{align}
   h^*_{\mu\nu}(x)&=-2t_{\mu\nu}(x)+4\lambda l^{-2}\int d^dy\sqrt{\hat\gamma(y)}G_{\mu\nu\eta\xi}(x,y)\Big(\delta^{\eta}_{\alpha}\delta^{\xi}_{\beta}-\frac{1}{d-2}\hat\gamma^{\eta\xi}(y)\hat\gamma_{\alpha\beta}(y)\Big)\Big[M^{\alpha\beta\rho\sigma}t_{\rho\sigma}\notag\\
&\quad+\frac{l^{2}}{4}\Big(4\hat\nabla_{\rho}t^{\alpha\beta}\hat\nabla_{\sigma}t^{\rho\sigma}-8\hat\nabla_{\rho}t^{\alpha\rho}\hat\nabla_{\sigma}t^{\beta\sigma}-2t^{\alpha\beta}\hat\nabla_{\sigma}\hat\nabla_{\rho}t^{\rho\sigma}+8t^{\beta\rho}\hat\nabla_{\sigma}\hat\nabla_{\rho} t^{\alpha\sigma}+4t^{\rho\sigma}\hat\nabla_{\sigma}\hat\nabla_{\rho}t^{\alpha\beta}+2t^{\alpha\beta}\hat\Box t^{\rho}_{\rho}\notag\\
&\quad-8t^{\beta\rho}\hat\Box t^{\alpha}_{\rho}+2t^{\rho}_{\rho}\hat\Box t^{\alpha\beta}-8t^{\alpha\rho}(\hat\nabla_{\rho}\hat\nabla_{\sigma}t^{\beta\sigma}+\hat\nabla_{\sigma}\hat\nabla_{\rho}t^{\beta\sigma}-\hat\Box t^{\beta}_{\rho}-\hat\nabla_{\sigma}\hat\nabla^{\beta}t_{\rho}^{\sigma})-4t^{\rho}_{\rho}\hat\nabla_{\sigma}\hat\nabla^{\beta}t^{\alpha\sigma}-\hat\gamma^{\alpha\beta}\hat\nabla_{\sigma}t^{\delta}_{\delta}\hat\nabla^{\sigma}t^{\rho}_{\rho}\notag\\
&\quad+8\hat\nabla_{\sigma}t^{\beta}_{\rho}\hat\nabla^{\sigma}t^{\alpha\rho}-4\hat\gamma^{\alpha\beta}\hat\nabla_{\rho}t^{\rho\sigma}\hat\nabla_{\delta}t_{\sigma}^{\delta}+4\hat\gamma^{\alpha\beta}\hat\nabla^{\sigma}t^{\rho}_{\rho}\hat\nabla_{\delta}t_{\sigma}^{\delta}+2\hat\gamma^{\alpha\beta}t^{\rho}_{\rho}\hat\nabla_{\delta}\hat\nabla_{\sigma}t^{\sigma\delta}-2\hat\gamma^{\alpha\beta}t^{\rho}_{\rho}\hat\Box t^{\sigma}_{\sigma}-2\hat\gamma^{\alpha\beta}\hat\nabla_{\sigma}t_{\rho\delta}\hat\nabla^{\delta}t^{\rho\sigma}\notag\\
&\quad+3\hat\gamma^{\alpha\beta}\hat\nabla_{\delta}t_{\rho\sigma}\hat\nabla^{\delta}t^{\rho\sigma}+4\hat\nabla_{\rho}t^{\beta\rho}\hat\nabla^{\alpha}t^{\sigma}_{\sigma}+4t^{\rho}_{\rho}\hat\nabla^{\alpha}\hat\nabla_{\sigma}t^{\beta\sigma}-4t^{\rho\sigma}\hat\nabla^{\alpha}\hat\nabla^{\beta}t_{\rho\sigma}-8\hat\nabla^{\sigma}t^{\alpha\rho}\hat\nabla^{\beta}t_{\rho\sigma}+4\hat\nabla^{\alpha}t^{\rho\sigma}\hat\nabla^{\beta}t_{\rho\sigma}\notag\\
&\quad+4\hat\nabla_{\rho}t^{\alpha\rho}\hat\nabla^{\beta}t^{\sigma}_{\sigma}-2\hat\nabla^{\alpha}t^{\rho}_{\rho}\hat\nabla^{\beta}t^{\sigma}_{\sigma}-2\hat\nabla_{\rho}t^{\sigma}_{\sigma}(\hat\nabla^{\rho}t^{\alpha\beta}-2\hat\nabla^{\beta}t^{\alpha\rho})-8\hat\nabla_{\sigma}t_{\rho}^{\sigma}\hat\nabla^{\beta}t^{\alpha\rho}-4t^{\rho}_{\rho}\hat\nabla^{\beta}\hat\nabla_{\sigma}t^{\alpha\sigma}+2t^{\rho}_{\rho}\hat\nabla^{\beta}\hat\nabla^{\alpha}t^{\sigma}_{\sigma}\Big)\Big](y)\notag\\
&\quad+O(\lambda^3),
\end{align}
where
\begin{align}
    t_{\mu\nu}(x)=\frac{1}{l^2}\int d^dx'\sqrt{\hat\gamma(x')}G_{\mu\nu\rho\sigma}(x,x')\Big[\hat T^{\rho\sigma}-\frac{1}{d-2}\hat T^{\alpha}_{\alpha}\hat\gamma^{\rho\sigma}\Big](x').
\end{align}
The corresponding stress tensor deformation is
\begin{align}
&\quad S^{(\lambda)}_{[2]}[\hat\gamma,\psi]\notag\\&=-\frac{\lambda^2l^{2}}{2}\int d^dx\sqrt{\hat\gamma}\Big[\frac{2}{l^2}t_{\mu\nu}t_{\rho\sigma}M^{\mu\nu\rho\sigma}+(t^{\rho}_{\rho})^2(\hat\nabla_{\nu}\hat\nabla_{\mu}t^{\mu\nu}-\hat\Box t^{\mu}_{\mu})-2t^{\mu\nu}(3\hat\nabla_{\mu}t^{\rho\sigma}\hat\nabla_{\nu}t_{\rho\sigma}-\hat\nabla_{\mu}t^{\rho}_{\rho}\hat\nabla_{\nu}t^{\sigma}_{\sigma}+4\hat\nabla_{\nu}t^{\sigma}_{\sigma}\hat\nabla_{\rho}t_{\mu}^{\rho}\notag\\
&\quad+4\hat\nabla_{\nu}t_{\mu}^{\rho}\hat\nabla_{\rho}t^{\sigma}_{\sigma}+4t_{\mu}^{\rho}\hat\nabla_{\rho}\hat\nabla_{\nu}t^{\sigma}_{\sigma}-4t_{\mu}^{\rho}\hat\nabla_{\rho}\hat\nabla_{\sigma}t_{\nu}^{\sigma}-2\hat\nabla_{\rho}t^{\sigma}_{\sigma}\hat\nabla^{\rho}t_{\mu\nu}-4\hat\nabla_{\rho}t_{\mu}^{\rho}\hat\nabla_{\sigma}t_{\nu}^{\sigma}-8\hat\nabla_{\nu}t_{\mu}^{\rho}\hat\nabla_{\sigma}t_{\rho}^{\sigma}+4\hat\nabla^{\rho}t_{\mu\nu}\hat\nabla_{\sigma}t_{\rho}^{\sigma}-4t^{\rho\sigma}\hat\nabla_{\sigma}\hat\nabla_{\nu}t_{\mu\rho}\notag\\
&\quad+4t^{\rho\sigma}\hat\nabla_{\sigma}\hat\nabla_{\rho}t_{\mu\nu}-4t_{\mu}^{\rho}\hat\nabla_{\sigma}\hat\nabla_{\rho}t_{\nu}^{\sigma}+t_{\mu\nu}\hat\nabla_{\sigma}\hat\nabla_{\rho}t^{\rho\sigma}+4t_{\mu}^{\rho}\hat\Box t_{\nu\rho}-t_{\mu\nu}\hat\Box t^{\rho}_{\rho}-4\hat\nabla_{\nu}t_{\rho\sigma}\hat\nabla^{\sigma}t_{\mu}^{\rho}-2\hat\nabla_{\rho}t_{\nu\sigma}\hat\nabla^{\sigma}t_{\mu}^{\rho}+6\hat\nabla_{\sigma}t_{\nu\rho}\hat\nabla^{\sigma}t_{\mu}^{\rho})\notag\\
&\quad-t^{\mu}_{\mu}(\hat\nabla_{\rho}t^{\sigma}_{\sigma}\hat\nabla^{\rho}t^{\nu}_{\nu}+4\hat\nabla_{\nu}t^{\nu\rho}\hat\nabla_{\sigma}t_{\rho}^{\sigma}-4\hat\nabla^{\rho}t^{\nu}_{\nu}\hat\nabla_{\sigma}t_{\rho}^{\sigma}-4t^{\nu\rho}(\hat\nabla_{\rho}\hat\nabla_{\nu}t^{\sigma}_{\sigma}-\hat\nabla_{\rho}\hat\nabla_{\sigma}t_{\nu}^{\sigma}-\hat\nabla_{\sigma}\hat\nabla_{\rho}t_{\nu}^{\sigma}+\hat\Box t_{\nu\rho})+2\hat\nabla_{\rho}t_{\nu\sigma}\hat\nabla^{\sigma}t^{\nu\rho}\notag\\
&\quad-3\hat\nabla_{\sigma}t_{\nu\rho}\hat\nabla^{\sigma}t^{\nu\rho})\Big].
\end{align}

The perturbation method can be employed to compute stress tensor deformations associated with more general higher-order gravitational actions. An important example is the Einstein gravity coupled with the Gauss-Bonnet term,
\begin{align}
    S_{G}&=S_{\text{EH}}+\alpha S_{\text{GB}}\notag\\
    &=l^{2}\int d^dx\sqrt{g}\Big[R+\alpha l^{2}(R^2-4R^{\mu\nu}R_{\mu\nu}+R^{\mu\nu\rho\sigma}R_{\mu\nu\rho\sigma})\Big].
\end{align}
For a saddle point satisfying condition $h^*=O(\lambda^0)$ to exist, the background metric must satisfy the Einstein equation
\begin{align}
    \hat G_{\mu\nu}+2\alpha l^2(\hat R\hat R_{\mu\nu}-2\hat R_{\mu\alpha}\hat R^{\alpha}{}_{\nu}-2\hat R^{\alpha\beta}\hat R_{\mu\alpha\nu\beta}+\hat R_{\mu}{}^{\alpha\beta\gamma}\hat R_{\nu\alpha\beta\gamma}-\frac{1}{4}\hat{\mathcal{L}}_{\text{GB}}\hat\gamma_{\mu\nu})=0.
\end{align}
The simplest choice is the flat background metric $\gamma_{\mu\nu}=\eta_{\mu\nu}$. Expanding the Gauss-Bonnet term in powers of $h_{\mu\nu}$, we have
\begin{align}
    S_{\text{GB}}=l^{4-D}\sum_{k=0}^{\infty}\int d^Dx\sqrt{\eta}\,\mathcal{R}^{(k)}_{\text{GB}},
\end{align}
with
    \begin{align}
\mathcal{R}^{(0)}_{\text{GB}}&=\mathcal{R}^{(1)}_{\text{GB}}=0,\notag\\
\mathcal{R}^{(2)}_{\text{GB}}&=\frac {\lambda^2}{2}(8\partial_{\mu}\partial^{\rho}h^{\mu\nu}\partial_{\rho}\partial_{\nu}h-4\partial_{\nu}\partial_{\mu}h\partial^2 h^{\mu\nu}-4\partial_{\nu}\partial_{\mu}h^{\mu\nu}\partial^2 h-4\partial_{\mu}\partial^{\rho}h^{\mu\nu}\partial_{\sigma}\partial_{\nu}h_{\rho}^{\sigma}\notag\\
   &\quad+2\partial_{\nu}\partial_{\mu}h^{\mu\nu}\partial_{\sigma}\partial_{\rho}h^{\rho\sigma}+2\partial_{\nu}\partial_{\mu}h_{\rho\sigma}\partial^{\sigma}\partial^{\rho}h^{\mu\nu}-4\partial_{\mu}\partial^{\rho}h^{\mu\nu}\partial_{\sigma}\partial_{\rho}h_{\nu}^{\sigma}+8\partial_{\mu}\partial^{\rho}h^{\mu\nu}\partial^2 h_{\nu\rho}\notag\\
    &\quad- 2\partial_{\sigma}\partial_{\nu}h_{\mu\rho}\partial^{\sigma}\partial^{\rho}h^{\mu\nu}-2\partial_{\rho}\partial_{\nu}h\partial^{\rho}\partial^{\nu}h+2\partial^2h\partial^2h-2\partial_{\nu}\partial_{\rho}h_{\mu\sigma}\partial^{\sigma}\partial^{\rho}h^{\mu\nu}\notag\\
    &\quad-2\partial^2 h^{\mu\nu}\partial^2h_{\mu\nu}+2\partial_{\sigma}\partial_{\rho}h_{\mu\nu}\partial^{\sigma}\partial^{\rho}h^{\mu\nu}).
\end{align}
When the background metric is flat, $\mathcal{R}^{(2)}_{\text{GB}}$ is a total derivative term. Therefore, the inclusion of the Gauss-Bonnet term does not change the first-order form of the stress tensor deformation~(\ref{firstleadingdeformation}).

\vspace{0.5cm}
\begin{center}{\large 
\textbf{
B.  $f(\mathcal{R})$ gravity in Palatini formalism}}
\end{center}
\vspace{0.15cm}
In this Supplemental Material, we briefly review some basics of Palatini formalism of $f(\mathcal{R})$ gravity. Varying with respect to \(g^{\mu\nu}\) and \(\Gamma^{\sigma}{}_{\mu\nu}\) yields EOM:
\begin{subequations}
\begin{align}
          f'(\mathcal{R})\mathcal{R}_{(\mu\nu)}-\frac{1}{2}f(\mathcal{R})g_{\mu\nu}&=-\lambda\hat T_{\mu\nu},\label{Palatini EoM}\\
    \bar\nabla_{\sigma}(\sqrt{g}f'(\mathcal{R})g^{\mu\nu})&=0.\label{Palatini EoM 2}
\end{align}
\end{subequations}
Here $\bar\nabla$ is the covariant derivative with respect to $\Gamma^{\sigma}{}_{\mu\nu}$. Taking the trace of the first EOM yields
\begin{align}
    f'(\mathcal{R})\mathcal{R}-\frac{d}{2}f(\mathcal{R})=-\lambda\hat T.
\end{align}
For a certain $f$, this equation is an algebraic equation in $\mathcal{R}$. The solution can be formally written as $\mathcal{R}=\mathcal{R}(\hat T(g^{*}))$. Plugging it into the gravitational action, we have
\begin{align}
    S^{(\lambda)}_{\text{EDA}}[g^{*},\psi]&=\hat S[g^{*},\psi]+\frac{1}{2\lambda}\int d^dx\sqrt{g^*}f(\mathcal{R}(\hat T(g^{*}))).
\end{align}
To obtain the specific form of the deformed action, we ought to find the exact solution $g^{*}$. By introducing the following metric conformal to $g_{\mu\nu}$ \cite{Sotiriou:2008rp},
\begin{align}
   \tilde g_{\mu\nu}&=(f'(\mathcal{R}))^{\frac{2}{d-2}}g_{\mu\nu},
\end{align}
the EOM (\ref{Palatini EoM 2}) can be rewritten as
\begin{align}
    \bar\nabla_{\sigma}(\sqrt{\tilde g}\ \tilde g^{\mu\nu})=0.
\end{align}
This equation is the definition of the Levi-Civita connection of $\tilde g_{\mu\nu}$, which gives
\begin{align}
    \Gamma^{\sigma}{}_{\mu\nu}=\frac{1}{2}\tilde g^{\sigma\lambda}(\partial_{\mu}\tilde g_{\nu\lambda}+\partial_{\nu}\tilde g_{\mu\lambda}-\partial_{\lambda}\tilde g_{\mu\nu}).
\end{align}
The Ricci curvature tensor after the conformal transformation is
\begin{align}
        \mathcal{R}_{\mu\nu} =R_{\mu\nu}-\frac{1}{f'(\mathcal{R})}\Big(\nabla_{\mu}\nabla_{\nu}+\frac{1}{d-2}g_{\mu\nu}\Box\Big)f'(\mathcal{R})+\frac{d-1}{d-2}\frac{1}{f'(\mathcal{R})^2}\nabla_{\mu}f'(\mathcal{R})\nabla_{\nu}f'(\mathcal{R}).
\end{align}
Plugging this expression into the EOM (\ref{Palatini EoM}), we obtain Einstein's field equation with a modified stress tensor,
\begin{align}
    &\quad R_{\mu\nu}-\frac{1}{2}Rg_{\mu\nu}\notag\\
    &=-\frac{\lambda}{f'}\hat T_{\mu\nu}-\frac{1}{2}(\mathcal{R}-\frac{f}{f'})g_{\mu\nu}+\frac{1}{f'}(\nabla_\mu\nabla_{\nu}-g_{\mu\nu}\Box)f'-\frac{d-1}{d-2}\frac{1}{f'^2}(\nabla_{\mu}f'\nabla_{\nu}f'-\frac{1}{2}g_{\mu\nu}\nabla^{\rho}f'\nabla_{\rho}f'),
\end{align}
which can be solved using the perturbation method in Supplemental Material A.

\vspace{0.5cm}
\begin{center}{\large 
\textbf{
C.  {Geometric realization of generalized Nambu-Goto action}}}
\end{center}
\vspace{0.15cm}
{ A notable example is the $d$-dimensional generalized Nambu-Goto action for a self-interacting scalar field,
\begin{align}
    &\quad S^{(\lambda)}_{\text{EDA}} = \int  d^dx\sqrt{g}\Big[\frac{1 - 2\lambda V - \sqrt{1 - 2\lambda (1 - \lambda V)\nabla^{\mu}\phi\nabla_{\mu}\phi}}{\lambda (1 - \lambda V)}\Big],\label{Effective deformed action of generalized Nambu-Goto action}
\end{align}
where $V=V(\phi)$. Such an effective deformed action satisfies the flow equation \cite{Ferko:2024zth, Babaei-Aghbolagh:2024hti}
\begin{align}
    \partial_{\lambda}S^{(\lambda)}_{\text{EDA}}&= \int  d^dx\sqrt{g}\Big[\frac{1}{2d}\text{tr}(T^2_{\text{EDA}}) - \frac{1}{d^2}(\text{tr}T_{\text{EDA}})^2\notag\\
    &\quad-\frac{d - 2}{2d^{3/2}\sqrt{d - 1}}\text{tr}T_{\text{EDA}}\sqrt{\text{tr}(T^2_{\text{EDA}}) - \frac{1}{d}(\text{tr}T_{\text{EDA}})^2}\Big].\label{Flow equation of generalized Nambu-Goto action}
\end{align}
From the perspective of gravity, the following action in the Palatini formalism (for $d\geq 3$) is postulated that could reproduce the generalized Nambu-Goto action,
\begin{align}
     \mathcal{A}^{(\lambda)}&=\frac{2 (d-1)(1-\lambda  V)\nabla^{\mu}\phi\nabla_{\mu}\phi}{f(V)}+\frac{f(V)}{4\lambda(d-1)}\Big(\frac{l^2\mathcal{R}}{(d-2)^2}-\frac{2l\sqrt{\mathcal{R}}}{1-\lambda  V}\Big)\notag\\
    &\quad+\frac{d-2}{4\lambda(1-\lambda V)}\Big(\frac{f(V)}{1-\lambda V}-2(1-2\lambda V)\Big),\label{gravitational action of the generalized NG action}
\end{align}
where the function $f(V)=d(1-\lambda V)(1-2\lambda V)-\sqrt{(1-\lambda V)^2(4-4d+d^2(1-2\lambda V)^2)}$. 
{  Ghost freedom requires $(1-\lambda V(\phi_0))f(V(\phi_0))>0$ and $\lambda f(V(\phi_0))>0$ at the stationary point $V'(\phi_0)=0$; in the Palatini form, the scalar is non-propagating, and no ghost arises. 
The effective cosmological constant is $\Lambda_{\mathrm{eff}}=(\lambda/2\ell^{2})[V(\phi_0)/\phi_0]$, whose sign depends on $V(\phi_0)$; both positive- and negative-curvature backgrounds are consistent provided these positivity conditions are satisfied.}
When $V(\phi)$ is constant, the last term of equation~(\ref{gravitational action of the generalized NG action}) shows that the deformation generates a volume term, yielding a cosmological constant as an intrinsic vacuum contribution.
By utilizing the EoM in the main text, the curvature terms present in (\ref{gravitational action of the generalized NG action}) are eliminated, thereby recovering the generalized Nambu-Goto action (\ref{Effective deformed action of generalized Nambu-Goto action}). 
In the special case $V(\phi)=0$, the action (\ref{gravitational action of the generalized NG action}) reduces to a minimally coupled free scalar with Palatini gravity, 
\begin{align}\label{naivemodel}
    \mathcal{A}^{(\lambda)}=(d-1)\nabla^{\mu}\phi\nabla_{\mu}\phi+\frac{1}{\lambda(d-1)}\Big(\frac{l^2\mathcal{R}}{2(d-2)^2}-l\sqrt{\mathcal{R}}\Big).
\end{align}}\par
Inspired by the deformed field theory action (\ref{Effective deformed action of generalized Nambu-Goto action}), we can more generally express the action in the following form,
\begin{align}
S^{(\lambda)}_{\text{EDA}}[g,\phi]&=\int d^dx\sqrt{g}\Big[\frac{1-\lambda A^{(\lambda)}-\sqrt{1-\lambda B^{(\lambda)}-2\lambda C^{(\lambda)}(\nabla^{\mu}\phi\nabla_{\mu}\phi)^{2q}}}{\lambda C^{(\lambda)}}\Big].\label{A30 effective deformed action}
\end{align}
The corresponding gravitational action is constructed as follows,
\begin{align}
    \mathcal{A}^{(\lambda)}=\frac{2q(d-2q)(\nabla^{\mu}\phi\nabla_{\mu}\phi)^{2q}}{F^{(\lambda)}(A,B)}-\frac{F^{(\lambda)}(A,B)}{4\lambda q^2(d-2q)}\Big(\mathcal{R}^q-\frac{qC^{(\lambda)}}{d-4q}\mathcal{R}^{2q}-\frac{(d-2q)(d-4q)[F^{(\lambda)}(A,B)-4q^2(1-\lambda A^{(\lambda)})]}{4qC^{(\lambda)}}\Big),\label{A31 action}
\end{align}
where $F^{(\lambda)}(A,B)=dq(1-\lambda A^{(\lambda)})-|q|\sqrt{(d-4q)^2-\lambda d^2A^{(\lambda)}(2-\lambda A^{(\lambda)})+8\lambda q(d-2q)B^{(\lambda)}}$. Here we set $l=1$ for simplicity. By using the EOM (12a) in the main text, we find
\begin{align}
    \mathcal{R}^q=\frac{(d-2q)(d-4q)[F^{(\lambda)}(A,B)-4q^2\sqrt{1-\lambda B^{(\lambda)}-2\lambda C^{(\lambda)}(\nabla^{\mu}\phi\nabla_{\mu}\phi)^{2q}}]}{2qC^{(\lambda)}F^{(\lambda)}(A,B)}.
\end{align}
Plugging it into (\ref{A31 action}), we recover the effective deformed action (\ref{A30 effective deformed action}). Some of the gravitational actions discussed in the text can be derived from (\ref{A31 action}). If we set $q=1/2$, $A^{(\lambda)}=2V$, $B^{(\lambda)}=0$, and $C^{(\lambda)}=(1-\lambda V)$, then (\ref{A31 action}) reduces to the gravitational action (\ref{gravitational action of the generalized NG action}), which corresponds to the Nambu-Goto action. If we set $q=1$, and $B^{(\lambda)}=A^{(\lambda)}(2-\lambda A^{(\lambda)})$, then (\ref{A31 action}) reduces to
\begin{align}
   \mathcal{A}^{(\lambda)}=\frac{(d-2)(\nabla^{\mu}\phi\nabla_{\mu}\phi)^{2}}{2(1-\lambda A^{(\lambda)})}-\frac{1-\lambda A^{(\lambda)}}{\lambda (d-2)}\Big(\mathcal{R}-\frac{C^{(\lambda)}}{d-4}\mathcal{R}^{2}\Big),
\end{align}
which corresponds to the quadratic Palatini $f(\mathcal{R})$ gravity with $\Lambda=0$, and $\alpha=-\frac{C^{(\lambda)}}{d-4}$. Suppose that $A^{(\lambda)}$ is a local function of $C^{(\lambda)}$ and has the Taylor expansion
\begin{align}
    A(C)=A_0+A_1C+A_2C^2+\cdots.
\end{align}
By taking the limit $C\to 0$, we obtain the Einstein's gravity,
\begin{align}
    \mathcal{A}^{(\lambda)}=\frac{(d-2)(\nabla^{\mu}\phi\nabla_{\mu}\phi)^{2}}{2(1-\lambda A_0)}-\frac{1-\lambda A_0}{\lambda (d-2)}\mathcal{R}.
\end{align}
The effective deformed action (\ref{A30 effective deformed action}) will remain finite in this limit if we impose the condition $1-\lambda A_0> 0$.

\vspace{0.5cm}
\begin{center}{\large 
\textbf{
D.  {Details of $T\bar T$-like deformation in $d$ dimensions}}}
\end{center}
\vspace{0.15cm}
In this Supplemental Material, we provide the computational details for the emergence of gravity from the $d$-dimensional $T\bar{T}$-like deformed field theory discussed in the main text. The effective deformed action can be formally written as $S^{(\lambda)}_{\text{EDA}}[g,\psi]=\int d^dx\sqrt{g}\mathcal{B}^{(\lambda)}(g^{\mu\nu},X_{\mu\nu},\psi)$, where $X_{\mu\nu}=\sum_{i,j}G_{ij}(\psi)\nabla_{\mu}\phi^{(i)}\nabla_{\nu}\phi^{(j)}$. The flow equation is given by
 \begin{align}
    \partial_{\lambda}S^{(\lambda)}_{\text{EDA}}&=\int d^dx\sqrt{g}\partial_{\lambda}\mathcal{B}^{(\lambda)}.\label{non-minimal flow equation 1}
\end{align}
The Lagrangian for the $d$-dimensional $T\bar{T}$-like deformation discussed in the main text is explicitly given by (14). By employing the expression of the eigenvalues of the effective stress-tensor $\tau_j^{(\lambda)}=2\chi_j\partial_{\chi_j}\mathcal{B}^{(\lambda)}-\mathcal{B}^{(\lambda)}$, we obtain
\begin{align}
    \tau^{(\lambda)}_j&=\lambda^{1-\Sigma}l^{\Delta}\frac{\beta_j^{\frac{p_j}{2}}}{\chi_j^{\frac{p_j}{2}}-\beta_j^{\frac{p_j}{2}}}\prod_{k=1}^{d}(\chi_k^{\frac{p_k}{2}}-\beta_k^{\frac{p_k}{2}})^{\frac{1}{p_k}}.
\end{align}
It follows that
\begin{align}
    \mathcal{B}^{(\lambda)}=\mathcal{B}_0+\lambda l^{\frac{\Delta}{1-\Sigma}} b^{\frac{1}{2(1-\Sigma)}}\Big(\prod_{j=1}^{d}(\tau_j^{(\lambda)})^{\frac{1}{p_j}}\Big)^{\frac{1}{\Sigma-1}},\label{non-minimal effective deformed action 1}
\end{align}
where $b=\prod_{j=1}^{d}\beta_j$. By combining (\ref{non-minimal flow equation 1})(\ref{non-minimal effective deformed action 1}), we obtain the flow equation (15) in the main text.\par
Next, we construct a class of gravitational actions capable of generating the effective deformed action given in (\ref{non-minimal effective deformed action 1}). For simplicity, we assume that the gravitational Lagrangian $\mathcal{A}^{(\lambda)}$ is a local function of the invariants $X_n$ and of those constructed from the Ricci curvature tensor, $\mathcal{R}_n = g^{\mu\nu}(\mathcal{R}^n)_{\mu\nu}$, respectively. The EOM (12a) in the main text reduces to
\begin{align}
2\sum_{m=1}^{d}m(\mathcal{R}^m)^{\mu}_{\nu}\partial_{\mathcal{R}_m}\mathcal{A}^{(\lambda)}+2\sum_{n=1}^{d}n(X^n)^{\mu}_{\nu}\partial_{X_n}\mathcal{A}^{(\lambda)}-\mathcal{A}^{(\lambda)}\delta^{\mu}_{\nu}=0.
\end{align}
Assuming that $\mathcal{R}^{\mu}_{\nu}$ can be diagonalized as $\mathcal{R}=\tilde U\text{diag}(\tilde r_1,\tilde r_2,...,\tilde r_d)\tilde U^{-1}$, we have
\begin{align}
&U^{-1}\tilde U\text{diag}(2\tilde r_1\partial_{\tilde r_1}\mathcal{A}^{(\lambda)},...,2\tilde r_d\partial_{\tilde r_d}\mathcal{A}^{(\lambda)})(U^{-1}\tilde U)^{-1}=\text{diag}(\mathcal{A}^{(\lambda)}-2\chi_1\partial_{\chi_1}\mathcal{A}^{(\lambda)},...,\mathcal{A}^{(\lambda)}-2\chi_d\partial_{\chi_d}\mathcal{A}^{(\lambda)}).
\end{align}
The diagonal elements of two matrices are identical, although the ordering of these elements differs. We can relabel the eigenvalues of $\mathcal{R}^{\mu}_{\nu}$ as $\lbrace{r_j}\rbrace$, which satisfy the differential equations
\begin{align}
    2r_j\partial_{r_j}\mathcal{A}^{(\lambda)}=\mathcal{A}^{(\lambda)}-2\chi_{j}\partial_{\chi_j}\mathcal{A}^{(\lambda)},\ \ \text{for}\ \ j=1,...,d.
\end{align}
The explicit form of $\mathcal{A}^{(\lambda)}$ is given by (16) in the main text. The EOM of the Ricci curvature tensor is
\begin{align}
    r_j^{\mathfrak{q}_j}=-\frac{\mathfrak{p}_j}{\mathfrak{s}_j}\chi_j^{\mathfrak{q}_j}+\frac{\beta_j}{\mathfrak{q}_j\mathfrak{s_j}},\ \ \ \ \ (\mathfrak{q}_j\neq 1).\label{non-minimal EoM R}
\end{align}
 Plugging (\ref{non-minimal EoM R}) into (16) in the main text, we can recover the effective deformed action (\ref{non-minimal effective deformed action 1}).\par
{  In the following, we employ the above formulas to analyze a specific example in two dimensions. The gravitational action is explicitly formulated as 
\begin{align}
    &\quad S_{\text{grav}}^{(\lambda)}[g,\Gamma,\psi]\notag\\
    &=\int  d^2x\sqrt{g}\Bigg\lbrace \mathcal{B}_0 + \frac{1}{\lambda l^{3}} \prod_{\theta=\pm} \bigg[ \sqrt{ \frac{l}{2} \Big( X + \theta \sqrt{2X^{\mu}_{\nu}X^{\nu}_{\mu} - X^2}}\Big)\notag\\
    &-  \sqrt{ M^2l +  \sqrt{ \mathcal{R} + \theta \sqrt{2\mathcal{R}^{\mu}_{\nu}\mathcal{R}^{\nu}_{\mu} - \mathcal{R}^2}} +  \sqrt{ X + \theta \sqrt{2X^{\mu}_{\nu}X^{\nu}_{\mu} - X^2}}}\bigg]  \Bigg\rbrace. \label{D=2 gravitational action}
\end{align}
Setting $p_1=p_2=1$, $\mathfrak{p}_1=\mathfrak{p}_2=\mathfrak{s}_1=\mathfrak{s}_2=l^{-1}$, $\mathfrak{q}_1=\mathfrak{q}_2=\frac{1}{2}$, and $\beta_1=\beta_2=-M^2/\sqrt{2}$, the action (16) in the main text can be identified with (\ref{D=2 gravitational action}). By solving the EOM (12a) in the main text, the specific form of the Ricci curvature tensor $\mathcal{R}^{\mu}_{\nu}$ is obtained.
\begin{align}
    \mathcal{R}^{\mu}_{\nu}&=(1+\frac{2\sqrt{2}M^2l}{\sqrt{\chi_+}+\sqrt{\chi_-}})X^{\mu}_{\nu}+\sqrt{2}M^2l(\sqrt{2}M^2l+\sqrt{\chi_+}+\sqrt{\chi_-}-\frac{X}{\sqrt{\chi_+}+\sqrt{\chi_-}})\delta^{\mu}_{\nu},
\end{align}
where $\chi_{\pm}=\frac{1}{2}(X\pm\sqrt{2X^{\mu}_{\nu}X^{\nu}_{\mu}-X^2})$.  Plugging it into the gravitational action (\ref{D=2 gravitational action}), and using the definition of the flow equation (\ref{non-minimal flow equation 1}), we obtain
\begin{align}
    \partial_{\lambda}S^{(\lambda)}_{\text{EDA}} = \frac{\sqrt{2}}{2M^2l^2} \int  d^2x\sqrt{g}\Big[(T_{\text{EDA}})^{\mu}_{\nu}(T_{\text{EDA}})^{\nu}_{\mu} - (T_{\text{EDA}})^2\Big],
\end{align}
which corresponds to the flow equation of the $T\bar T$ deformation.}

\vspace{0.5cm}
\begin{center}{\large 
\textbf{
E.  {Stress-tensor two-point function of free scalar field theory in four dimensions}}}
\end{center}
\vspace{0.15cm}
{ To derive the first-order deformed partition function induced by a generic quadratic non-local stress-tensor deformation, one should evaluate the precise form of the stress-tensor two-point function in a curved background. In this Supplementary Material, we perform explicit calculations in the case of four-dimensional free scalar field theory. The one-loop effective action \cite{Vassilevich:2003xt} is given by
\begin{align}
    W[g]=\frac{1}{2}\text{ln }\text{det}(-\Box+m^2),
\end{align}
where $m$ is the mass of the scalar field. By employing zeta-function regularization, the functional determinant admits a representation via the heat kernel associated with the operator $(-\Box+m^2)$,
\begin{align}
    W[g]=-\frac{1}{2}\int^{\infty}_{0}\frac{d\tau}{\tau}\text{tr}(e^{-\tau(-\Box+m^2)}).
\end{align}
For $m \neq 0$, the integral converges in the infrared limit $\tau \to \infty$, while it diverges in the ultraviolet limit $\tau \to 0^+$, with the divergence structure controlled by the leading terms of the heat kernel expansion. We employ the proper-time regularization by introducing the UV cutoff at $\tau=\Lambda^{-2}$. From the heat kernel expansion, the regularized one-loop effective action can be expressed as
\begin{align}
    W_{\Lambda}[g]&=-\frac{1}{2}\int^{\infty}_{\Lambda^{-2}}\frac{d\tau}{\tau}\text{tr}(e^{-\tau(-\Box+m^2)})\notag\\
    &=-\frac{1}{64\pi^2}\int d^4x\sqrt{g}\Big[\Lambda^4+\frac{\Lambda^2}{3}\Big(R-6m^2\Big)-\frac{2\text{ln}(\Lambda/m)}{3}\Big(m^2R-3m^4-\frac{1}{5}\Box R-\frac{1}{12}R^2\notag\\
    &\quad+\frac{1}{30}R_{\mu\nu}R^{\mu\nu}-\frac{1}{30}R_{\mu\nu\rho\sigma}R^{\mu\nu\rho\sigma}\Big)\Big]+O(\Lambda^0).
\end{align}
The stress-tensor two-point function can be calculated by variational principle
\begin{align}\label{2pt general}
    \langle{T^{\mu\nu}(x)T^{\rho\sigma}(y)}\rangle&=\frac{4}{\sqrt{g(x)}\sqrt{g(y)}}\frac{\delta^2 W_{\Lambda}}{\delta g_{\mu\nu}(x)\delta g_{\rho\sigma}(y)}\notag\\
    &=-\frac{1}{64\pi^2}\Big[(\Lambda^4-2m^2\Lambda^2+2m^4\text{ln}(\Lambda/m))(g^{\mu\nu}g^{\rho\sigma}-g^{\mu\rho}g^{\nu\sigma}-g^{\mu\sigma}g^{\nu\rho})\delta(x-y)\notag\\
    &\quad+\frac{2}{3}(\Lambda^2-2m^2\text{ln}(\Lambda/m))\Big((2g^{\mu\nu}g^{\rho\sigma}+g^{\mu\rho}g^{\nu\sigma}+g^{\mu\sigma}g^{\nu\rho})\Box-2g^{\mu\nu}\nabla^{\rho}\nabla^{\sigma}-2g^{\rho\sigma}\nabla^{\mu}\nabla^{\nu}\notag\\
    &\quad-4R^{\mu\rho\nu\sigma}+2g^{\mu\nu}R^{\rho\sigma}+2g^{\rho\sigma}R^{\mu\nu}-g^{\mu\rho}R^{\nu\sigma}-g^{\mu\sigma}R^{\nu\rho}-g^{\nu\rho}R^{\mu\sigma}-g^{\nu\sigma}R^{\mu\rho}\Big)\delta(x-y)\notag\\
    &\quad+\text{ln}(\Lambda/m)\mathcal{F}_{(4)}(x,y)\Big]+O(\Lambda^0),
\end{align}
where $\mathcal{F}_{(4)}(x,y)$ involves fourth-order derivatives of the $\delta$-function, including $\nabla^4\delta$, $R\nabla^2\delta$, $R^2\delta$, etc.\par
Next, we examine the first-order deformed partition function induced by a non-local quadratic stress-tensor deformation,
\begin{align}
    \mathcal{Z}^{(\lambda)}=\mathcal{Z}_0\Big(1-\frac{\lambda}{2l^2}\int d^4xd^4y\sqrt{g(x)g(y)}H_{\mu\nu\rho\sigma}(x,y)\langle{T^{\mu\nu}(x)T^{\rho\sigma}(y)}\rangle\Big).
\end{align}
Here $H_{\mu\nu\rho\sigma}(x,y)$ represents a general kernel. An illustrate example is the kernel $H_{\mu\nu\rho\sigma}(x,y)=G_{\mu\nu\rho\sigma}(x,y)-\frac{1}{2}G_{\mu\nu\alpha\beta}(x,y)g^{\alpha\beta}(y)g_{\rho\sigma}(y)$ derived from the linearized Einstein gravity calculation in the main text, where $G_{\mu\nu\rho\sigma}(x,y)$ satisfies the differential equation $[-g^{\mu\rho}g^{\nu\sigma}\Box-R^{\mu\rho\nu\sigma}-R^{\nu\rho\mu\sigma}]_xG_{\rho\sigma\alpha\beta}(x,y)=\delta(x-y)\delta^{\mu}_{\alpha}\delta^{\nu}_{\beta}$. The Green's function admits a representation via the heat kernel of the corresponding Laplacian-type operator,
\begin{align}
    G_{\mu\nu\rho\sigma}^{(\Lambda)}(x,y)=\int_{\Lambda^{-2}}^{\infty}d\tau\ K_{\mu\nu\rho\sigma}(\tau;x,y).
\end{align}
In four dimensions, the heat kernel has the following small-$\tau$ expansion,
\begin{align}\label{off-diag K}
    K_{\mu\nu\rho\sigma}(\tau;x,y)=\frac{\Delta^{1/2}(x,y)}{16\pi^2\tau^2}e^{-\sigma(x,y)/2\tau}\sum_{n=0}^{\infty}a^{(2n)}_{\mu\nu\rho\sigma}(x,y)\tau^n,
\end{align}
where $\sigma$ is the Synge world function and $\Delta$ is the Van Vleck–Morette determinant. The trace of the heat kernel can be written in terms of the geometric invariants,
\begin{align}
     K_{\mu\nu\rho\sigma}(\tau;x,x)=\frac{1}{16\pi^2}\Big[\frac{1}{2\tau^2}(g_{\mu\rho}g_{\nu\sigma}+g_{\mu\sigma}g_{\nu\rho})-\frac{1}{\tau}\Big(R_{\mu\rho\nu\sigma}+R_{\nu\rho\mu\sigma}-\frac{1}{12}R(g_{\mu\rho}g_{\nu\sigma}+g_{\mu\sigma}g_{\nu\rho})\Big)+O(\tau^0)\Big].
\end{align}
Combining the stress-tensor two-point function (\ref{2pt general}) and the heat kernel (\ref{off-diag K}), and employing the coincidence limit $x\to y$ of derivatives of the Synge world function and the Van Vleck–Morette determinant \cite{Avramidi:2000bm,Avramidi:2001ns,Vassilevich:2003xt}, we compute the first-order correction to the partition function in a systematic expansion,
\begin{align}\label{A78}
\mathcal{Z}^{(\lambda)}_{\text{reg}}&=\mathcal{Z}_0+\frac{\lambda\mathcal{Z}_0}{128\pi^2l^2}\int d^4x\sqrt{g}\Big[-\frac{1}{4\pi^2}\Big(13\Lambda^6-10m^2\Lambda^4-16m^2\Lambda^4\text{ln}(\Lambda/m)+10m^4\Lambda^2\text{ln}(\Lambda/m)\Big)\notag\\
&\quad-\frac{R}{12\pi^2}\Big(7\Lambda^4+8\Lambda^4\text{ln}(\Lambda/m)-14m^2\Lambda^2\text{ln}(\Lambda/m)+16m^4(\text{ln}(\Lambda/m))^2\Big)+O(\nabla^4)\Big]\notag\\
&=\mathcal{Z}_0+\frac{\lambda\mathcal{Z}_0}{2l^2}\int d^{4}x\sqrt{g}\,
\Big[
\alpha_0+\alpha_{1} R
+O(\nabla^4)
\Big],
\end{align}
where $O(\nabla^4)$ represents the contribution of higher-derivative gravitational terms.\par
From the above computation, the one-loop correction can be written in the form of an effective action,
\begin{align}
    S_{\text{eff}}&=\frac{\lambda}{128\pi^2l^2}\int d^4x\sqrt{g}\Big[\frac{1}{4\pi^2}\Big(13\Lambda^6-10m^2\Lambda^4-16m^2\Lambda^4\text{ln}(\Lambda/m)+10m^4\Lambda^2\text{ln}(\Lambda/m)\Big)\notag\\
&\quad+\frac{R}{12\pi^2}\Big(7\Lambda^4+8\Lambda^4\text{ln}(\Lambda/m)-14m^2\Lambda^2\text{ln}(\Lambda/m)+16m^4(\text{ln}(\Lambda/m))^2\Big)+O(\nabla^4)\Big].
\end{align}
Next, counterterms are introduced to cancel the divergent parts, leading to the renormalized action. Note that both the constant term and the Einstein-Hilbert term contain power divergences in $\Lambda$, which are directly removed by counterterms. Moreover, the Einstein-Hilbert term contains a purely logarithmic divergence, $16m^4(\ln(\Lambda/m))^2$. Upon introducing the renormalization scale $\mu$, this logarithmic term can be rewritten as 
\begin{align}
    16m^4(\ln(\Lambda/m))^2=16m^4\Big((\ln(\Lambda/\mu))^2+2\ln(\Lambda/\mu)\ln(\mu/m)+(\ln(\mu/m))^2\Big),
\end{align}
where the last term contributes to the finite part after renormalization. Thus, the renormalized effective action takes the form
\begin{align}
    S_{\text{ren}}(\mu)&=\frac{\lambda}{128\pi^2l^2}\int d^4x\sqrt{g}\Big[\frac{4m^4}{3\pi^2}(\text{ln}(\mu/m))^2R+O(\nabla^4)\Big]\notag\\
    &=\frac{1}{16\pi G_{\text{eff}}(\mu)}\int d^4x\sqrt{g}\Big[R+O(\nabla^4)\Big].
\end{align}

}

\end{document}